\documentclass[10pt,twocolumn,aps,pra,showpacs,superscriptaddress,floatfix]{revtex4-1}

\usepackage[colorlinks=true,citecolor=blue,urlcolor=blue]{hyperref}
\usepackage[utf8]{inputenc}
\usepackage[T1]{fontenc}
\usepackage[sc,osf]{mathpazo}
\usepackage{amsmath, amsthm, amssymb,amsfonts}
\usepackage{graphicx}
\usepackage{dcolumn}
\usepackage{bm}
\usepackage{bbm}
\usepackage{mathtools}
\usepackage{comment}
\usepackage{color}

\usepackage{graphicx}
\usepackage{amssymb}
\usepackage{amsmath}
\usepackage{amsthm}
\usepackage{color}
\usepackage{psfrag}
\usepackage{epsfig}
\usepackage{bbm}
\usepackage{bm}
\hypersetup{breaklinks}

\newcommand{\beq}{\begin{eqnarray}}
\newcommand{\eeq}{\end{eqnarray}}

\newtheorem{theorem}{Theorem}
\newtheorem*{theorem*}{Theorem}

\newtheorem{obs}{Observation}

\begin{document}
\title{Indistinguishability of causal relations from limited marginals}
\author{Costantino Budroni}
\affiliation{Naturwissenschaftlich-Technische Fakult\"at, Universit\"at Siegen, Walter-Flex-Str. 3, 57068 Siegen, Germany}
\author{Nikolai Miklin}
\affiliation{Naturwissenschaftlich-Technische Fakult\"at, Universit\"at Siegen, Walter-Flex-Str. 3, 57068 Siegen, Germany}
\author{Rafael Chaves}
\affiliation{International Institute of Physics, Federal University of Rio Grande do Norte, 59070-405 Natal, Brazil}
\affiliation{Institute for Theoretical Physics, University of Cologne, 50937 Cologne, Germany}

\date{\today}
\begin{abstract}
We investigate the possibility of distinguishing among different causal relations starting from 
a limited set of marginals. Our main tool is the notion of adhesivity, that is, the extension of 
probability or entropies defined only on subsets of variables, which provides additional 
independence constraints among them. Our results provide a criterion for recognizing which causal structures are indistinguishable when only limited marginal information is accessible. Furthermore, the existence of such extensions greatly simplifies the characterization of a marginal scenario, a result that facilitates the derivation of Bell inequalities both in the probabilistic and entropic frameworks, and the identification of marginal scenarios where classical, quantum, and postquantum probabilities coincide.
\end{abstract}
\maketitle

\section{Introduction}

Deciding global properties of a given object from partial information is a 
problem often encountered in the most diverse fields. Just to cite a few examples, one can 
mention: knowledge integration of expert systems \cite{lauritzen1988}, database theory 
\cite{studeny1994marginal}, causal discovery \cite{Pearlbook,Spirtesbook} and the phenomenon of 
quantum nonlocality \cite{Bell1964,Brunner2014review}. More formally, in all such applications 
we are facing the so-called marginal problem: deciding whether a given set of marginal 
probability distributions for some random variables arises from a joint distribution of all 
these variables. Naturally, extensions to the quantum realm are also known: the quantum 
marginal problem \cite{Klyachko2006} asks whether marginal density operators describing 
physical systems have a global extension, a fundamental question in quantum information with 
the most diverse applications \cite{Schilling2013,Walter2013}.

The reasons for our partial and/or marginal knowledge of a given system will intrinsically depend on 
the context and may have a variety of fundamental or practical reasons. For instance, in 
quantum mechanics incompatible observables cannot be perfectly and jointly measured. In turn, 
artificial intelligence systems gather information from several sources, each providing partial 
(typically overlapping) information about the global system \cite{Vomlel1999}.

In this paper, we will be mainly interested in the marginal problem arising in causal 
discovery \cite{Pearlbook,Spirtesbook} where the aim is to decide, based on empirical observations, which underlying causal structures can explain our data. In particular, we provide a general result connecting the limited information available, i.e., the marginals, and the set of causal structures that can be distinguished on the basis of such information.

Importantly, causal discovery includes the test of local hidden variable (LHV) models that play a fundamental role in the foundations of quantum mechanics, in particular in Bell's theorem \cite{Bell1964} and its practical applications in the processing of information \cite{Brunner2014review}. 
Not surprisingly, given the importance and wide breadth of applications of the marginal problem within causal inference \cite{Pearlbook,Spirtesbook} and related fields \cite{Bonet2001,friedman2004inferring,Steeg2011,steudel2015information}, several approaches have been formulated in order to solve it. In the following, we briefly describe some of the leading approaches.

In the absence of hidden variables ---that is, all variables composing a given causal structure 
are empirically available--- causal discovery can be faithfully performed based on the set of 
conditional independencies (CIs) implied by the model under test \cite{Pearlbook,Spirtesbook}. 
However, causal models with hidden variables imply highly non-trivial constraints on the level 
of the observed distributions \cite{Pearl1995,Geiger1999,Tian2002,Garcia2005}, constituting 
still a very active field of research \cite{Kang2012b,lee2015causal,Chaves2016poly,Rosset2016}. 
In fact, from the causal perspective, Bell's theorem can be understood as a particular class of 
a causal discovery problem, where the underlying causal assumptions are those of local 
causality and measurement independence \cite{Spekkens2012,Chaves2015b}. Furthermore, as 
realized by Pitowsky \cite{Pitowsky1989,Pitowsky1991}, the set of probability distributions 
compatible with such causal assumptions defines a convex set, more precisely a polytope, which 
facets are exactly the famous Bell inequalities. 

The problem with this probabilistic approach, the most used in the study of Bell nonlocality, 
is that its useful linear convexity property (allowing the derivation of Bell inequalities via efficient 
linear programs) does not generalize to more complicated causal structures \cite{Garcia2005}. 
In the general case, one has to resort to algebraic geometry tools 
\cite{hartshorne2013algebraic} that, at least in principle, are able to solve the marginal 
problem for arbitrary causal structures \cite{Geiger1999}. However, in practice, because of its 
high computational complexity, its use in causal inference is restricted to very few cases of 
interest \cite{Garcia2005,lee2015causal}. Thus, alternative approaches have also been pursued.

Instead of looking at the constraints imposed on the level of probabilities, 
Braunstein and Caves \cite{Braunstein1988} asked what are the LHV constraints on the level of 
the Shannon entropies of these probabilities, which directly lead to the concept of entropic 
Bell inequalities. Apart from its fundamental relevance from a information-theoretic point of 
view 
\cite{Pawlowski2009,Barnum2010entropy,Short2010,Safi2011,janzing2013,Chaves2015device,Janzing2015algorithmic}, 
the entropic approach stands as a meaningful alternative for the fact that it provides a much 
more convenient route for the study of complex causal network beyond LHV models 
\cite{Chaves2014,Chaves2014b,Chaves2016entropic} and for extensions of causal discovery to the 
case of quantum causal structures 
\cite{Fritz2012,chaves2014information,Henson2014,weilenmann2016entropy,weilenmann2016non,pienaar2016causal}.

The entropic approach to the marginal problem and causal discovery has received growing attention 
\cite{Cerf1997,Chaves2012,FritzChaves2013,Chaves2013a,Kurzy2012,Kurzy2014,raeisi2015entropic,wajs2015information,Devi2013,rastegin2014generalized,rastegin2014formulation,cao2016experimental}, 
but still suffers from two main drawbacks. The first stands from a computational complexity 
issue. The region of compatible entropies characterizing a given causal structure form a convex set, thus enormously simplifying the problem as compared to the highly nontrivial nonconvex sets appearing in the algebraic geometry approach. In spite of that, the derivation of entropic Bell inequalities mostly relies on the elimination of variables from a system of linear inequalities via the so-called Fourier-Motzkin elimination algorithm \cite{Williams1986} that has a double-exponential computational complexity, which limits its use to very few cases of interest. The second issue arises from the fact that this approach relies on an outer approximation of the region defining valid entropies for a collection of variables \cite{Yeung2008}. Finding better approximations to the entropy cone is a very active field of research in both classical \cite{Yeung2008} and quantum information theory 
\cite{Linden2005,Cadney2012}, but to our knowledge very few results \cite{weilenmann2016non} 
are known about the implications of it for marginal problems, that is, the projection of the 
entropy cone on the subspace defined by the empirically observable variables.

Within this context, the goal of this paper is to propose a way for characterizing the correlations compatible with a given causal structure. With that aim, we work out the consequences of the the so-called adhesivity property \cite{vorobev1967coal,vorobev1963markov,matus2007} in the context of marginal scenarios, and what are its implications for the distinguishability between generic causal structures. Furthermore, by applying the general algorithm to particular cases of interest, we show two by-product advantages of our approach: i) it provides a faster computational algorithm and ii) in some cases provides a better approximation to the true marginal set of correlations characterizing a given causal model.

\section{Summary of the results}

Given the amount of preliminary notions needed to understand our main results and the length of the associated sections, it is convenient to first give an informal summary of such results.
The main problem we address is the possibility of distinguishing different causal structures associated with a set of random variables starting from limited information, i.e., limited marginals of their probability distribution. In Sec.~\ref{sec:ad_ind_con}, using the notion of adhesivity we identify which causal structures are always consistent with a given marginal scenario (Th.~\ref{th:extT}). These causal structures are defined as Markov random fields, starting from a graph-theoretic procedure (triangulation) applied to the marginal scenario (hyper)graph.

This result is then used in Sec.~\ref{sec:ind} to prove the main theorem (Th.~\ref{th:caus_and_adh}) on the relation between causal structures and marginal scenarios. In simple terms, Th.~\ref{th:caus_and_adh} identifies which causal structures can be falsified, i.e., proven to be inconsistent with a given set of marginals. This identification is done on the basis of the independence relations associated with a causal structure and the corresponding Markov random fields associated with the marginal scenario.

An immediate application of these results is in the characterization of which causal structures and marginals scenarios can lead to a different set of allowed correlations depending on the classical, quantum or even post-quantum nature of the underlying process; an important step in the generalization of Bell's theorem to more complex cases \cite{Henson2014,pienaar2016causal}
 and in the understanding of quantum correlations via informational principles such as information causality \cite{Pawlowski2009}. For instance, as a consequence of (Th.~\ref{th:extT}) it follows that if the marginal scenario corresponds to an acyclic hypergraph, every probability distribution is compatible with a classical description, thus precluding the possibility of observing quantum nonlocality in such cases. A similar conclusion holds for scenarios satisfying the condition in case i) of Th.~\ref{th:caus_and_adh}.

In addition, Th.~\ref{th:extT} is also used to improve the characterization and approximation of entropy cones and correlation polytopes associated with a marginal scenario, under no assumption of  a causal structure (Obs.~\ref{obs:cones_inclusion}), thus facilitating the derivation of new Bell inequalities. Th.~\ref{th:caus_and_adh} also takes into account these methods and explains when they can be applied to causal structures.

The paper is organized as follows. In Sec.~\ref{prenotions}, we provide a detailed survey of 
all concepts and tools required for the understanding of the paper. More precisely, in 
Sec.~\ref{sec:g_and_h}, we describe hypergraphs and their properties; in \ref{sec:c_s}, we 
have a brief account of causal structures; in \ref{sec:marg}, we introduce the notion of a 
marginal scenario and finally in Sects.~\ref{sec:corrpoly} and \ref{sec:e_c}, we review the 
tools provided by correlation polytopes and entropic cones, respectively. 
In Sec.~\ref{sec:ad_ind_con}, we start describing some of the original results in this paper, 
namely, which causal relations are always consistent with a given marginal scenario. In 
Sec.~\ref{sec:opt_ent}, we show how the notion of adhesivity can be used to compute the Bell 
inequalities of a given marginal scenario using the associated minimal hypergraph, which in 
some cases can greatly reduce the computational complexity of the problem. In 
Sec.~\ref{sec:ind}, we prove general results concerning the indistinguishability of causal 
structures while in 
Sec.~\ref{sec:ex}, we put our general approach to analyze a few cases of interest. In 
Sec.~\ref{sec:concl}, we summarize our findings and discuss interesting future directions of research.
\section{Preliminary notions}
\label{prenotions}

\subsection{Graphs and hypergraphs}\label{sec:g_and_h}
\begin{figure}[t]
    \centering
    \begin{minipage}[b]{0.2\textwidth}
        \includegraphics[width=\textwidth]{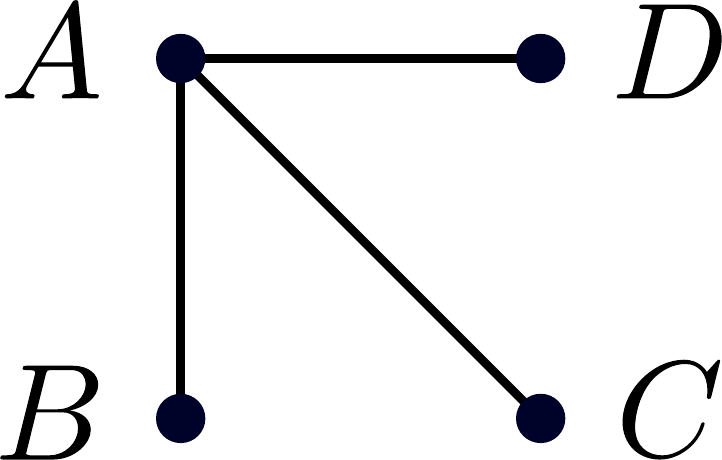}
        {(a) $\mathcal{T}$}
    \end{minipage}
    \hspace{0.02\textwidth}
    ~ 
    \begin{minipage}[b]{0.2\textwidth}
    \includegraphics[width=\textwidth]{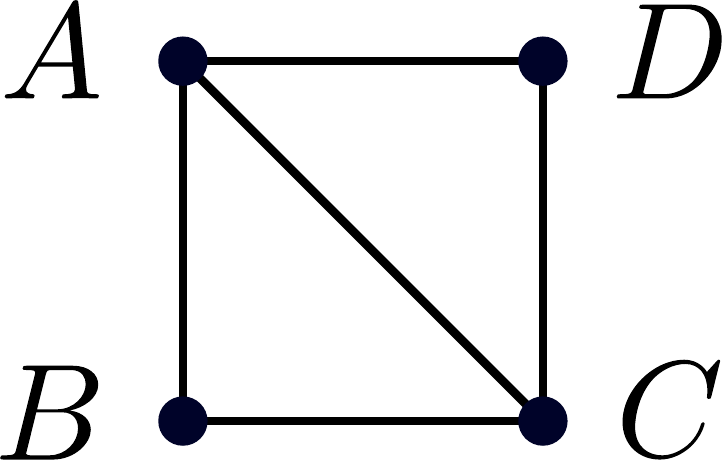}
    {(b) $\mathcal{G}$}
    \end{minipage}\\[12pt]
    \begin{minipage}[b]{0.2\textwidth}
    \includegraphics[width=\textwidth]{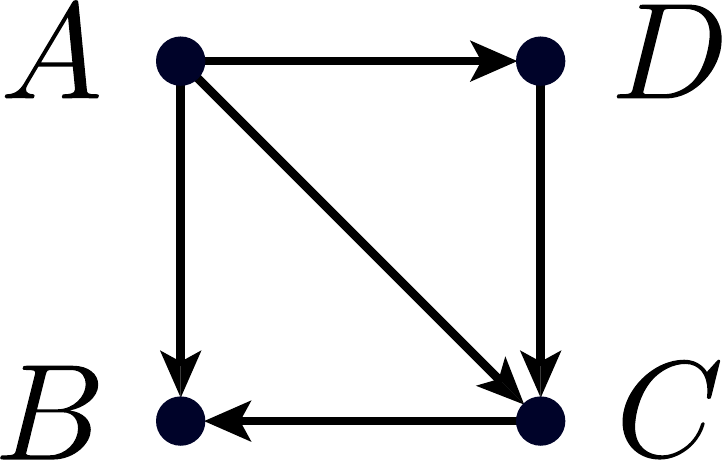}
    {(c) DAG}
    \end{minipage}
    \hspace{0.02\textwidth}
     \begin{minipage}[b]{0.2\textwidth}
    \includegraphics[width=\textwidth]{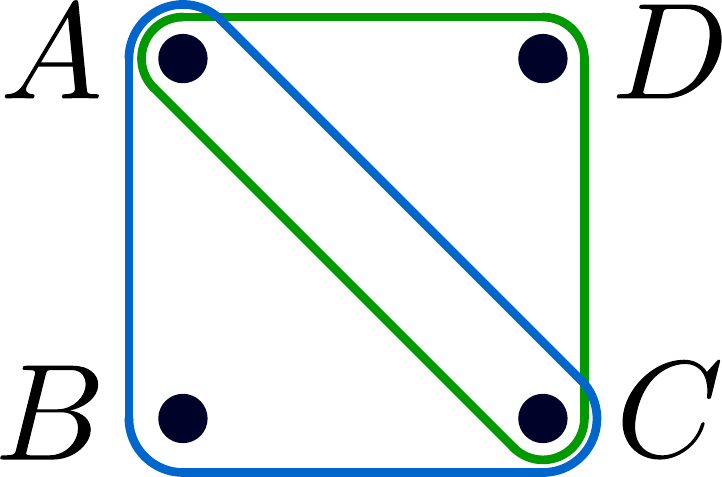}
    {(d) $\mathcal{H}$}
    \end{minipage}
    
\caption{\label{fig:2a}\textbf{Examples of graphs and hypergraphs}. (a) A tree graph $\mathcal{T}$ where $\{B,A,D\}$ is a path. (b) A graph $\mathcal{G}$ where $\{A,C,B,A\}$ is a loop and $\{A,B,C\}$ and $\{A,C,D\}$ are cliques. $B$ and $D$ are separated by $\{A,C\}$. 
(c) A directed graph where  $\{A,C,B,A\}$ is not a directed path, because direction of $(A,B)$ is not respected. This graph does not contain loops i.e. it is a directed acyclic graph (DAG). $B$ and $D$ are as well separated by $\{A,C\}$, but $\{C\}$ is a minimal separator. (d) A hypergraph $\mathcal{H}$ where nodes $B$ and $D$ are separated by $\{A,C\}$.
}
\end{figure}

A hypergraph $\mathcal{H}=(\mathcal{N},\mathcal{E})$ is defined by a finite set of nodes 
$\mathcal{N}=\{1,\ldots,n\}$ and a set of (hyper)edges corresponding to subsets of 
$\mathcal{N}$, i.e., $\mathcal{E}\subset 2^{\mathcal{N}}$.
A graph $\mathcal{G}$ is a special case of an hypergraph where edges have cardinality 2, i.e., 
$\mathcal{G}=(\mathcal{N},\mathcal{E})$, with $|E|=2$ for all $E\in \mathcal{E}$. A graph can 
also be directed, i.e., have directed edges corresponding to ordered pairs 
$(i,j)\in \mathcal{E}$, denoted by an arrow from $i$ to $j$. In the following, by graph we will 
always mean a undirected graph, unless stated otherwise. See Fig.~\ref{fig:2a} for examples of 
graphs, directed graphs, hypergraphs, and additional notions discussed below.

Since we will be interested only in hypergraphs without isolated nodes, we will assume that 
$\mathcal{N}=\cup_{E\in\mathcal{E}} E$, when not stated otherwise, and we will sometimes denote 
the hypergraph simply by the set of edges $\mathcal{E}$.

Paths, cycles, and acyclicity are fundamental notions in graph theory. A {\it path} is a sequence of distinct nodes $v_{0},\ldots,v_{n}$  (except possibly the first and last) connected by edges $(v_{k},v_{k+1})$ $k=0,\ldots,n$, and a {\it closed path} or a {\it loop} is a path with first and last node coinciding, i.e., $v_{0}=v_{n}$. For directed graphs, the definition is analogous with  $(v_{k},v_{k+1})$ representing a directed edge. {\it Acyclic graphs},  also called {\it tree graphs}, are graphs not containing loops. A graph is {\it connected} if, for every pair of nodes, there is a path connecting them.

\begin{figure}
    \centering
    \begin{minipage}[b]{0.2\textwidth}
        \includegraphics[width=\textwidth]{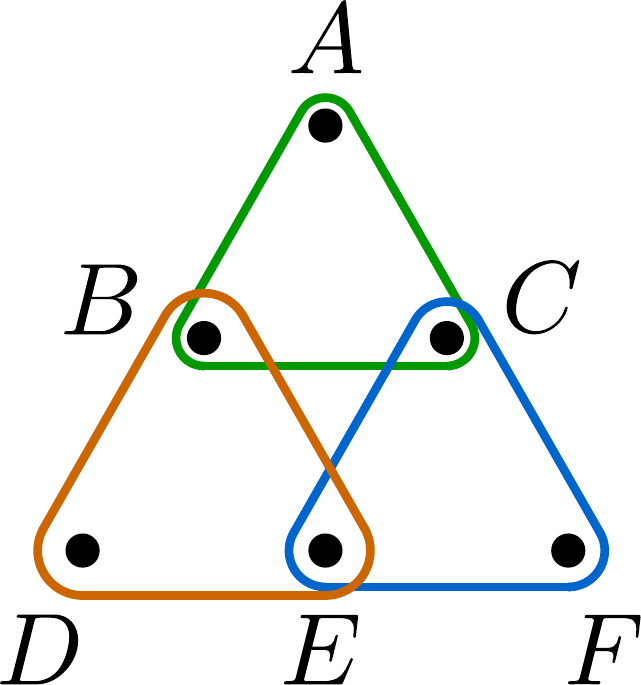}
         {(a) $\mathcal{H}$}
    \end{minipage}
    \hspace{0.02\textwidth}
    ~ 
    \begin{minipage}[b]{0.2\textwidth}
    \includegraphics[width=\textwidth]{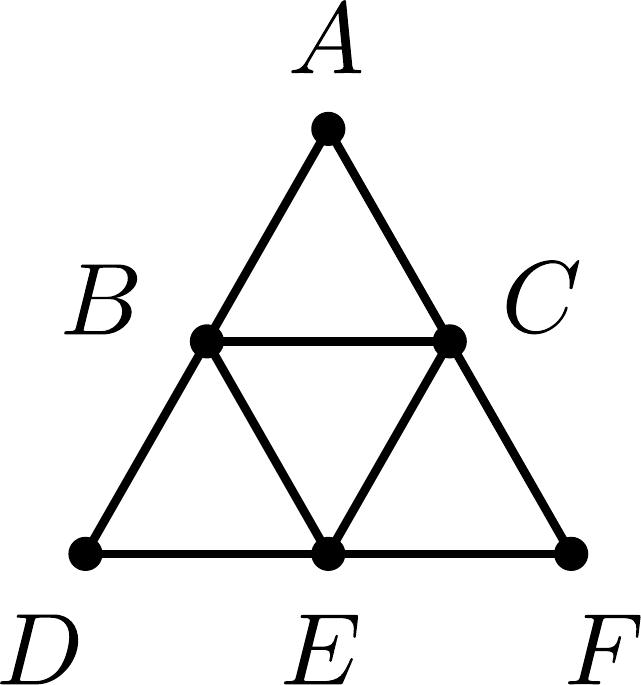}
     {(b) $\mathcal{G}$}
    \end{minipage}
    ~
    
    \begin{minipage}[b]{0.2\textwidth}
    \includegraphics[width=\textwidth]{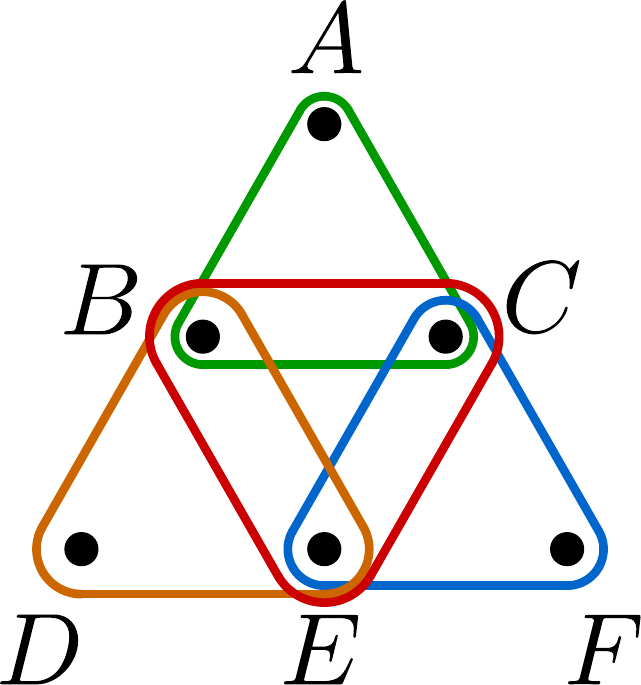}
     {(c) $\mathcal{H}'$ }
    \end{minipage}
    \caption{\textbf{Example of graphs and hypergraphs}. (a) A cyclic hypergraph $\mathcal{H}$. (b) A 2-section graph $\mathcal{G}=[\mathcal{H}]_2$ of the hypergraph. (c) A clique hypergraph $\mathcal{H}'$ of $\mathcal{G}$. Notice that $\mathcal{H}'\neq \mathcal{H}$. In fact, $\mathcal{H}'$  is an extension of $\mathcal{H}$.
     }\label{fig:acyclic}
\end{figure}

A {\it clique} is a set of nodes $v_0,\ldots,v_n$ pairwise connected by an edge, i.e. $(v_i,v_j)\in \mathcal{E}$ for all  $i,j=1,\ldots,n$, $i\neq j$. Given a graph $\mathcal{G}$, we can construct a hypergraph from it, called the {\it clique hypergraph} $\mathcal{H}_\mathcal{G}^{\rm cl}$, with the same nodes and hyperedges in $\mathcal{H}_\mathcal{G}^{\rm cl}$ corresponding to cliques in $\mathcal{G}$.
Similarly, a hypergraph $\mathcal{H}=(\mathcal{N},\mathcal{E})$ can be transformed into a graph by constructing the {\it 2-section} $[\mathcal{H}]_2$:  we connect by edges in $\mathcal{G}$ all nodes that are connected by at least one hyperedge in $\mathcal{H}$. Notice that given a hypergraph $\mathcal{H}$, the clique graph of its 2-section will have, in general, extra hyperedges with respect to $\mathcal{H}$ (cf. Fig.~\ref{fig:acyclic}).

A hypergraph $\mathcal{H}=(\mathcal{N},\mathcal{E})$ is a {\it partial hypergraph} of $\mathcal{H}'=(\mathcal{N},\mathcal{E}')$ if for any $E\in \mathcal{E}$ there exist $E' \in \mathcal{E}'$ such that $E\subset E'$. Equivalently, we will say that $\mathcal{H}'$ {\it 
extends}, or is an {\it extension} of, $\mathcal{H}$ (cf. Fig.~\ref{fig:acyclic}).

Given two disjoint subsets of nodes $A,B$ they are said to be {\it separated} by a subset $C$ 
if for each pair $a\in A, b\in B$, all the paths from $a$ to $b$ pass through $C$, i.e., if we 
remove $C$, $A$ and $B$ are no longer connected. In addition, $C$ is called a minimal separator 
if $C\diagdown \{v_i\}$ is no longer a separator for any $v_i\in C$.

An important notion is also that of {\it triangulated}, or {\it chordal} graphs, namely, graphs 
for which every cycle $v_{0},\ldots,v_{n}$ of length $n\geq 4$, contains a {\it chord}, i.e., 
an edge connecting $(v_i,v_{i+2})$. Given any graph, $\mathcal{G}$, additional edges can be added 
such that the obtained graph, $\mathcal{G}'$, is triangulated, and we will refer to
$\mathcal{G}'$ as the {\it triangulation} of $\mathcal{G}$, see, e.g., 
Fig.~\ref{fig:4a}~(b),(c). 

For hypergraphs, the generalization of the notions of acyclicity and tree is not 
straightforward and several definition have been proposed (cf. Ref.~\cite{beeri1983}). For 
reasons that will be clear in Sec.~\ref{sec:ad_ind_con}, here we will focus on the notion of 
$\alpha$-acyclicity, developed in the framework of database theory, which we will simply call 
acyclicity.  There are several equivalent characterizations of this property (cf. 
Refs.~\cite{lauritzen1996graphical,beeri1983}), but we will focus on three of them: a 
characterization via the so-called {\it Graham algorithm}, one via the {\it running 
intersection property} of hyperedges, and the characterization as a  clique hypergraph of a 
chordal graph.

The Graham algorithm is defined as follows. Given a hypergraph described by hyperedges 
$\mathcal{E} = \{E_1,\ldots,E_n\}$, apply the following operations whenever they are possible
\begin{itemize}
\item[$a)$] Delete a node $i$ if it appears in exactly one hyperedge.
\item[$b)$] Delete a hyperedge $E$ if $E\subset E'$.
\end{itemize}
Acyclic hypergraphs are those for which the Graham algorithm returns the empty set. Given a hypergraph $\mathcal{H}$, its {\it reduced hypergraph} is the hypergraph obtained by applying only operation $b)$ of the Graham algorithm.

A hypergraph has the running intersection property if there exists an ordering of the edges, $E_1,\ldots,E_n$ such that
\begin{equation}\label{eq:run}
 S_i := E_i \cap (E_1\cup \ldots\cup E_{i-1}) \subset E_j, \text{ with  } j< i.
\end{equation}
In addition, for a connected and reduced hypergraph, the set $\{S_i\}$ corresponds to the {\it set of 
minimal separators} of the graph, i.e., $S_i$ separates $R_i :=E_i\diagdown S_i$ 
from $(E_1\cup \ldots\cup E_{i-1})\diagdown S_i$. It can be proven that the running intersection property is equivalent to the empty set output 
for the Graham algorithm, so it can be used as an alternative definition of an acyclic hypergraph (see, e.g., \cite{lauritzen1996graphical}).

The third equivalent property is defined in terms of graphs: a hypergraph is acyclic iff its 
hyperedges correspond to the set of cliques of a triangulated graph (see e.g., 
Ref.~\cite{lauritzen1996graphical}).

In order to clarify the above notions, it is instructive to apply them to the simple example 
depicted in Fig.~\ref{fig:acyclic}. For instance, we can apply the Graham algorithm to the 
hypergraph in Fig.~\ref{fig:acyclic} (a). By applying operation $(a)$, we remove the nodes $A,D,F$ and 
we are left with the edges $\{ \{B,C\},\{B,E\},\{C,E\} \}$. At this point the algorithm stops, 
because we cannot remove any edge via operation $(b)$, or any other node with operation $(a)$. 
The hypergraph $\mathcal{H}$ is thus not acyclic. We can apply the same procedure 
to $\mathcal{H}'$: by removing the nodes $A,D,F$ we are left with the edges 
$\{ \{B,C,E\}, \{B,C\},\{B,E\},\{C,E\} \}$. We can then continue and remove the edges  
$\{B,C\},\{B,E\},\{C,E\}$ via operation $(b)$, and finally the nodes $B,E,C$ connected by a 
single edge $\{B,C,E\}$. The hypergraph $\mathcal{H}'$ is thus acyclic. Equivalently, one can 
see that the graph $\mathcal{G}$, obtained as the 2-section $[\mathcal{H}]_2=[\mathcal{H}']_2$, 
has as cliques exactly the hyperedges of $\mathcal{H}'$, but not those of $\mathcal{H}$. 
Finally, for the running intersection property of $\mathcal{H}'$, we can choose the ordering 
$E_1= \{B,C,E\}$, and for $E_2,E_3,E_4$ any ordering of the remaining edges. It is clear that 
any intersection of edges is contained in $E_1$. One can also straightforwardly check that 
$\mathcal{H}$ does not have the running intersection property. Similarly, one can easily check 
the property of separators of the sets $S_i =  E_i \cap (E_1\cup \ldots\cup E_{i-1})$, 
$i=2,3,4$, for the hypergraph $\mathcal{H}'$.

\subsection{Causal structures}\label{sec:c_s}

For a set of random variables $X_1, \dots, X_n$ some additional logical and/or causal constraints may apply. Usually, these constraints correspond to nonlinear constraints for probabilities (e.g., factorization properties) and to linear constraints for entropies (e.g., vanishing of the mutual information). Two common examples are given below: 
\begin{itemize}
    \item {\it Deterministic dependence:} The variable $X_i$ is said to be a deterministic function of $X_j$ if their joint probability distribution satisfies $P(x_i,x_j)=\delta_{x_i,F(x_j)} P(x_j)$, where the deterministic dependence is given by $x_i=F(x_j)$. Such types of constraints are usually present in network coding \cite{Yeung2008}. While not strictly necessary, deterministic constraints also play an important role in the derivation of Bell inequalities \cite{Fine1982}.
    \item {\it Conditional or unconditional independence:} The variable $X_i$ is said to be independent of the variable $X_j$ if the joint probability distribution satisfies $P(x_i,x_j)=P(x_i)P(x_j)$. Similarly, variable $X_i$ is said to be conditionally independent (CI) of $X_j$ given $X_k$, if $P(x_i,x_j,x_k)=P(x_i|x_k)P(x_j|x_k)P(x_k)$; that is, $X_k$ screens off the correlations between the two other variables. We will denote the two situations as $(X_i \perp X_j)$ and $(X_i \perp X_j | X_k)$, respectively.
\end{itemize}
Upper-case letters (e.g., $X$) denote the random variables, and lower-case letters (e.g., $x$) the specific value they assume.

As we will see next, conditional independence and unconditional independence play a crucial role in the study of Bayesian networks and Markov random fields \cite{lauritzen1996graphical}; this is why we will focus on them in the remaining of the paper.

\subsubsection{Bayesian networks}
A Bayesian network (BN) is a probabilistic model for which conditional dependencies can be represented via a directed acyclic hypergraph (DAG). More precisely, the probability distribution factorizes as
\begin{equation}\label{eq:dagfac}
P(x_1,\ldots,x_n)=\prod_{i=1}^n P(x_i| {\rm Pa}_i),
\end{equation}
where ${\rm Pa}_i$ denotes the {\it parents} of the node $i$, i.e., the nodes with arrows pointing at $i$. The above factorization of the probability distribution gives rise to the {\it local Markov property} 
\begin{equation}
(X_i \perp {\rm Nd}_i | {\rm Pa}_i),
\end{equation}
namely that $X_i$ is independent of its {\it nondescendants} ${\rm Nd}_i$, i.e., nodes reachable from $X_i$ via a directed path, given its parents. 

\begin{figure}[h!]
    \centering
    \begin{minipage}[b]{0.2\textwidth}
        \includegraphics[width=\textwidth]{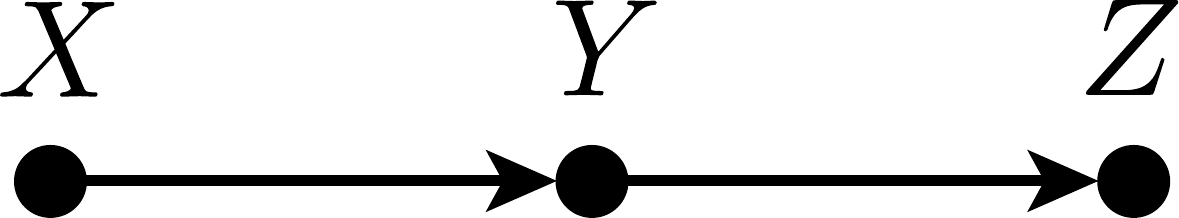}
         { (a) }
    \end{minipage}
    \hspace{0.02\textwidth}
        ~ 
    \begin{minipage}[b]{0.2\textwidth}
    \includegraphics[width=\textwidth]{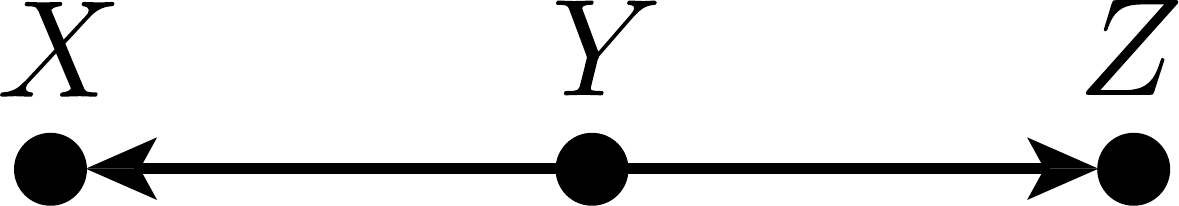}
     { (b) }
    \end{minipage}
    
    ~
    \begin{minipage}[b]{0.2\textwidth}
    \includegraphics[width=\textwidth]{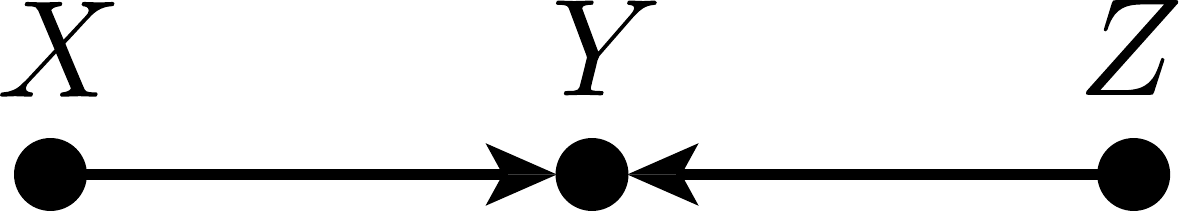}
     { (c) }
    \end{minipage}
    \caption{\textbf{Examples of different Bayesian networks}. (a) A DAG representing a Markov chain $X\rightarrow Y \rightarrow Z$ implying the CI $(X \perp Z | Y )$. (b) A DAG where the variable $Y$ is a common parent of $X$ and $Z$, once more implying the CI $(X \perp Z | Y )$. (c) A DAG where the variables $X$ and $Z$ have a common child $Y$. In this case $(X \perp Z)$, but $(X \not\perp Z | Y )$.}\label{fig:bayes}
\end{figure}

More generally, one has the set of conditional independence relations
\begin{equation}\label{eq:bayes_ind}
\mathcal{I}(\mathcal{G})=\{ (X_A \perp X_B | X_C )\ |\ {\rm dsep}_\mathcal{G}(A:B|C)\},
\end{equation}
where ${\rm dsep}_\mathcal{G}(A:B|C)$ refers to the $d$-separation properties of nodes in $A$ and $B$ with respect to nodes in $C$, namely that every path from $a\in A$ to $b\in B$, or vice versa, is {\it blocked} by a node in $C$. The path is said to be blocked if it contains one of the following: $x\rightarrow c \rightarrow y$, or $x\leftarrow c \leftarrow y$, or $x\rightarrow z \leftarrow y$, for $x,y,z,c$ in the path, $c\in C$, $z\notin C$ and no descendant of $z$ is in $C$ (cf. Ref.~\cite{Pearlbook}).

Bayesian networks are of particular relevance to formalize causal relations. Within this context, such causal models have been called causal Bayesian networks \cite{Pearlbook}, as opposed to traditional Bayesian networks that formalize conditional independence relations without having necessarily a causal interpretation. To exemplify, consider the three DAGs shown in Fig.~\ref{fig:bayes}. DAGs (a) and (b) clearly imply a different set of causal relations between variables $X$, $Y$ and $Z$: in both cases the correlations between $X$ and $Z$ are mediated via $Y$ but in (a) $X$ is a parent of $Y$ while in (b) the reverse is true. In spite of their clear causal differences, both causal models imply the same set of CIs, namely that $(X \perp Z | Y )$. That is, every observable probability distribution $p(x,y,z)$ compatible with (a) is also compatible with (b), thus both models are indistinguishable from observations alone \footnote{In such cases, in order to distinguish different causal structures, one has to rely on another crucial concept of the mathematical theory of causality, that of an intervention \cite{Pearlbook}}. The DAG (c) in Fig.~\ref{fig:bayes} can nonetheless be distinguished from (a) and (b), since it implies that $(X \not\perp Z | Y )$ and the only CI is given by the independence constraint $(X \perp Z)$.

\subsubsection{Markov random fields}
Similarly to Bayesian networks, Markov random fields (MRF) correspond to probabilistic models for which conditional dependencies can be represented by a graph $\mathcal{G}$. In this case, the graph is undirected and it may contain cycles. More precisely, independence relations are given by the {\it global Markov property}
\begin{equation}
(X_A \perp X_B | X_C )
\end{equation}
if every path from a node in $A$ to a node in $B$ passes through a node in $C$, i.e., if $C$ is a separator for $A$ and $B$ in $\mathcal{G}$, a fact denoted as ${\rm sep}_\mathcal{G}(A:B|C)$. We will denote the corresponding set of independence relations as
\begin{equation}\label{eq:mrf_ind}
\mathcal{I}(\mathcal{G})=\{ (X_A \perp X_B | X_C )\ |\ {\rm sep}_\mathcal{G}(A:B|C)\}
\end{equation}

As opposed to Bayesian networs (cf. Eq.~\eqref{eq:dagfac}), MRFs do not admit a unique factorization of the probability distribution. However, for the special case of a triangulated graph, denoting with $C_1,\ldots,C_k$ the set of maximal cliques with the running intersection property and $S_i:= C_i \cap (C_1\cup \ldots\cup C_{i-1})$, as in Eq.~\eqref{eq:run}, one can write
\begin{equation}\label{eq:mrffac}
P(x_1,\ldots,x_n) = \prod_{i=1}^k \frac{P(x_{C_i})}{P(x_{S_i})}.
\end{equation}

Bayesian networks can be described also via MRFs, via the so called {\it moral graph} (cf. Ref. \cite{lauritzen1996graphical}), however this comes at the price of losing some of the original independence constraints described by the DAG. 

\subsection{Marginal scenarios}\label{sec:marg}

In many relevant situations, one may have only partial information of the distribution of the variables. This may be due to practical limitations in collecting data, e.g., latent variables which cannot be measured, or fundamental limitation, e.g., the impossibility of performing a joint measurement of incompatible quantum observables. This is common in Bell and noncontextuality experiments \cite{Brunner2014review,thompson2013recent}, where one has access only to a limited set of joint probability distributions. Also in purely classical contexts the role of partial information can hardly be overemphasized \cite{Pearlbook,Steeg2011}. For instance, in the so called instrumentality tests modeling randomized experiments \cite{Pearlbook,Bonet2001}, the effects from a drug in the recovery of patients is allowed to depend on some unobserved factors that are not under experimental control (social or economical background, etc).

For this reason, we introduce the notion of a marginal scenario. Given a set of random variables $X=\{X_1,\ldots,X_n\}$, a {\it marginal scenario} is a collection of subsets $\mathcal{M}=\{ M_1,\ldots,M_{|\mathcal{M}|}\}$, $M_i\subset X$ of them representing variables that can be jointly measured, i.e., for each $M_i$, we have access to a probability distribution $P_{M_i}(x_{M_i})$. Moreover, if a set of variables are jointly measurable, we require that the same holds for any subset, i.e., $M\in \mathcal{M}$ and $M'\subset M$ imply $M'\in \mathcal{M}$. Equivalently, one can take only the maximal subsets $S\in \mathcal{M}$.

A marginal scenario can be naturally considered as an hypergraph, with $\mathcal{M}$ the set of hyperedges and $\cup_i M_i$ the set of nodes. We will adopt the maximal subsets convention above and assume that $S\in \mathcal{M}$ are only maximal subsets, i.e., the hypergraph is reduced. We will call such a hypergraph the {\it marginal scenario hypergraph}, or simply marginal scenario when it is clear we are referring to the hypergraph.

\begin{figure}
\includegraphics[width=0.3\textwidth]{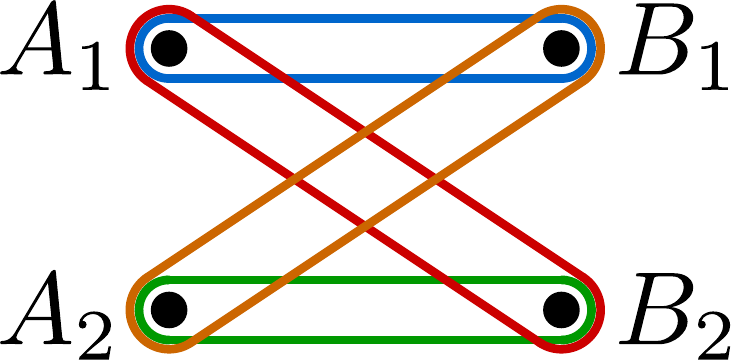}
\caption{Hypergraph of the marginal scenario associated with a Bell-CHSH experiment. The observed probabilities correspond to the marginal for $\{A_x,B_y\}$, for $x,y=1,2$.}\label{fig:chshhg}
\end{figure}

It is again instructive to consider a simple example to fix the above notions. A standard example is given by a Bell experiment \cite{Bell1964}, in particular by the Clauser-Horne-Shimony-Holt (CHSH) scenario \cite{Clauser1969}. Two parties, Alice and Bob, perform measurements on their part of a shared quantum system. They can perform one of two measurements, labeled as $A_1,A_2$ for Alice and $B_1,B_2$ for Bob. The observed probabilities will then amount to the marginals for $\{A_x,B_y\}$. The corresponding marginal scenario hypergraph is depicted in Fig.~\ref{fig:chshhg}.

\subsection{Correlation polytopes and non-local correlations}
\label{sec:corrpoly}
In quantum information, one of the applications of the concept of a marginal scenario is exactly in the study of Bell's theorem and quantum nonlocality \cite{Bell1964,Brunner2014review}. The simplest Bell scenario is given by two distant (ideally space-like separated) parties, Alice and Bob, that at each round of their experiment receive particles produced by a common source. Upon receiving their shares of the physical system, they measure a given observable of it recording the measurement outcome. The measurements choices by Alice and Bob (that are assumed to be independent of how the system has been prepared) are labeled by the variables $X$ and $Y$ while their outcomes are given by $A$ and $B$, respectively. Under the assumption of local realism, every observable probability distribution that can be obtained in such experiment can be decomposed as the so called local hidden variable (LHV) model 
\begin{equation}
\label{eq.LHV}
p(a_x,b_y)=p(a,b \vert x,y)= \sum_{\lambda} p(\lambda)p(a\vert x,\lambda)p(b\vert y,\lambda),
\end{equation}
where the variable $\lambda$ stands for a full description of the source producing the particles and any other mechanism that might affect the measurement outcomes.

As realized by Pitowsky \cite{Pitowsky1989}, the set of probability distributions compatible with the LHV model form a convex set of correlations, the so called correlation polytope. This polytope is characterized by finitely many extremal points, exactly those representing deterministic functions in the LHV decomposition \eqref{eq.LHV}. Equivalently, this polytope is characterized by finitely many linear inequalities, the non-trivial ones being exactly the Bell inequalities.

Different methods to characterize such correlation polytopes are available. Here, we will describe a particular method obtained via Fine's theorem \cite{Fine1982} and that plays a fundamental role in the derivation of entropic Bell inequalities as we will see in 
Sec.~\ref{sec:e_c}.

Fine's theorem shows that the existence of a LHV model of the form \eqref{eq.LHV} is equivalent to the existence of a joint probability distribution $p(a_1,\dots,a_m,b_1,\dots,b_m)$ that marginalizes to the observed distribution $p(a_x,b_y)$ with $x,y=1,\dots,m$. The existence of a well-defined joint distribution over all variables  implies that such distribution must respect some constraints, namely, positivity and normalization. That is, $p(a_1,\dots,a_m,b_1,\dots,b_m)$ must lie inside a simplex polytope \cite{boyd_convex_2009}. From this geometric perspective, the correlation polytope is nothing other than the projection of the simplex polytope --characterizing the joint distribution-- to a subspace of it that is given by the marginal scenario in question, that is, a projection to the subspace spanned by the observable components $p(a_x,b_y)$ with $x,y=1,\dots,m$. Such a projection can be obtained by eliminating, from the corresponding system of linear 
inequalities describing the simplex, all terms that correspond to non-observables probabilities. This can be achieved, for example, via a standard algorithm known as Fourier-Motzkin  
elimination \cite{Williams1986}. Once the redundant inequalities are removed, the remaining set 
gives the facets of the correlations polytope, that is, tight Bell inequalities.

Unfortunately, such a nice linear convex picture does not hold for more complicated causal structures \cite{Branciard2010,Fritz2012,Kang2012b,Chaves2016poly} that now require computationally expensive and highly intractable methods from algebraic geometry \cite{Geiger1999} in order to deal with the non-linear constraints arising from such models. As mentioned before and explained in more details in the next section, this is one of the reasons why the entropic approach has become a more viable option in the study of complex causal structures. In short, polynomial constraints on the level of probabilities are turned into simpler linear relations in terms of entropies.

\subsection{Entropic cone}
\label{sec:e_c}
Given a collection of $n$ discrete random variables $X_1, \dots, X_n$ with an associated joint probability distribution $P(x_1, \dots,x_n)$, and denoting with $X_S$ the random vector $(X_i)_{i\in S}$, for any subset $S\subset [n]:=\{1, \dots, n\}$, the Shannon entropy $H: 2^{[n]} \to \mathbbm{R}$ is defined as
\begin{equation}
H(S):= H(X_S) =-\sum_{x_S}P(x_S)\log_2 P(x_S).
\end{equation}

 The above entropies can be arranged in a vector ${h=(H(\emptyset),H(X_1),\ldots,H(X_{1},X_{2}),\ldots, H(X_1, \dots, X_n))}$ $\in \mathbbm{R}^{2^n}$. The region
\begin{equation}
\Gamma_{[n]}^* :=	\overline{\left\{ h \in \mathbbm{R}^{2^n} \,|\, h = (H(S))_{S\subset [n]} \text{ for some entropy
	} H \right\}},
\end{equation}
where the overline denotes the closure in $\mathbb{R}^{2^n}$, is known to be a convex cone, also called the {\it entropy cone} and it has been studied extensively in information theory \cite{Yeung2008}. A tight and explicit description, however, has not yet been found for $n>3$, but only some outer approximations of $\Gamma_{[n]}^*$ via polyhedral cones, i.e., cones described by a finite system of linear inequalities $Ax\geq b$, where $A$ is an $m\times n$ real matrix and $b$ an $m$-dimensional real vector.
The most famous outer approximation to the entropic cone is the so-called {\it Shannon cone} $\Gamma_{[n]}$, defined by
\begin{subequations}\label{shannonineqs_basic}
\begin{eqnarray}
	h_{}([n]\setminus\{i\}) &\leq& h_{}([n]),
	\\
	h_{}(S) + h_{}(S\cup\{i,j\}) &\leq& h_{}(S\cup \{i\}) + h_{}(S\cup \{j\}),\
	\\
	h_{}(\emptyset) &=& 0,
\end{eqnarray}
\end{subequations}
for all $i, j\in [n]$, $i \neq j$,   and $S \subset [n] \setminus\{i,j\}$. That is, the Shannon cone associated with $n$ variables is described by $2^{n-2}\binom{n}{2}+n$ inequalities plus one equality constraint (normalization).

The above is the minimal set of inequalities that implies monotonicity of entropy, i.e., $H(A|B)\geq 0$, and the submodularity (or 
strong subadditivity), i.e., $I(A:B|C):= H(A,C)+H(B,C)-H(A,B,C)-H(C)\geq 0$, for any disjoint subsets $A,B,C\subset [n]$ (cf. 
Ref.~\cite{Yeung2008}).

Inequalities in Eq.~\eqref{shannonineqs_basic} are known as {\it Shannon-type inequalities} in information theory or 
{\it polymatroidal axioms} in combinatorial optimization \cite{Yeung2008}. Given a finite set $N$ and real-valued function 
$f:2^N\rightarrow \mathbb{R}$, the pair $(N,f)$ is called a polymatroid if $f$ satisfies Eqs.~\eqref{shannonineqs_basic} above for 
$[n]=N$ and $S,\{i,j\}\subset N$.

Geometrically, the entropy cone $\Gamma_{\mathcal{M}}$ associated with the marginal scenario 
$\mathcal{M}$ corresponds to the projection of the
entropic cone onto the subspace of the corresponding variables. Similarly to what happens to the projection of a simplex polytope into a correlation polytope, for a polyhedral cone, such a 
projection can be obtained by eliminating, from the corresponding system of linear 
inequalities, all terms that correspond to non-observables terms. After removing redundant inequalities, the remaining set  gives facets of the entropic cone in the observable subspace.

Conditional independence constraints arising from BNs or MRFs, i.e., of the form 
$(X_A \perp X_B | X_C )$, corresponds to linear constraints on the vector of entropies, i.e., 
hyperplanes defined by the equation ${I(A:B|C)=0}$.
Linear constraints from causal structures not only reduce the dimension of the problem, but as 
well, when applied to polymatroids, reduce the number of axioms needed to describe the 
constrained cone. For instance, it was shown in Ref.~\cite{reducedaxioms} that if 
$I(A:B|C)=0$, with $A,B,C$ disjoint sets of variables, then from the set of polymatroid axioms 
in Eq.~\eqref{shannonineqs_basic}, the following inequalities are redundant
\begin{eqnarray}
I(A:E|BC)\geq 0, \\
I(B:E|AC)\geq 0,
\end{eqnarray}
where $E\in[n]\setminus\{A,B\}\cap C$. Hence, in this case one can generate a compact 
representation of polymatroid axioms \cite{reducedaxioms}.

\section{Adhesivity and independence constraints associated with a marginal scenario}\label{sec:ad_ind_con}
The main result of this section is that when one has only partial information about (i.e., only 
some marginals of) a probability distribution, such marginals are always consistent with a global distribution where additional independence constraints are imposed. We start by introducing the notion of adhesivity and restating in our language a theorem by Vorob'ev \cite{vorobev1967coal} (Th.~\ref{th:MRFext}), then we connect this result with the notion of marginal scenario to prove which independence constraints are always compatible with a set of marginals (Th.~\ref{th:extT}).

On the one hand, such additional constraints simplify the characterization of the entropy cone 
and correlation polytope associated with the marginal scenario. On the other hand, this result 
allows us to identify which causal structures 
can be distinguished when we have access only to some restricted set of marginals. 

The notion of adhesivity, albeit in different terms, was first introduced for probability 
distributions \cite{vorobev1967coal,vorobev1963markov}  and subsequently extended to entropies 
\cite{matus2007}. In the framework of Bell and noncontextuality inequalities, similar ideas have been investigated by several authors \cite{BudroniMorchio2010, budroni2012bell,Kurzy2012,budroni2012bell2,Chaves2014}, but never in full generality.

\subsection{Adhesivity of probabilities}\label{sec:ad}

Adhesivity can be explained in simple terms as follows. Given two sets of variables $X_I=(X_i)_{i\in I}$ 
and $X_J=(X_j)_{j\in J}$ and two probability distributions $p(x_I)$ and $p'(x_J)$  such that $p$ 
and $p'$ coincide on the variables $X_{I\cap J}$, we can define a probability distribution on 
$I\cup J$ as
\begin{equation}\label{eq:proext}
P(x_{I\cup J}) = \left\lbrace
\begin{array}{lc}
0 & \mbox{ if } p(x_{I\cap J}) = 0,\\
\frac{p(x_I)p'(x_J)}{p(x_{I\cap J})} & \mbox{ otherwise. }
\end{array} \right.
\end{equation}
One can easily check that this is a valid probability distribution on the set of variables in 
$I\cup J$.

The construction in Eq.~\eqref{eq:proext} implies that every two marginals of a probability 
distribution are always consistent with a probability distribution conditionally independent of 
their intersection, i.e., $(X_{I\diagdown J}\ \perp\ X_{J\diagdown I}\ |\ X_{I\cap J} )$ , 
since $p(x_I)p'(x_J)/p(x_{I\cap J})= p(x_I| x_{I\cap J}) p'(x_J |x_{I\cap J}) p(x_{I\cap J})$.
We call such an extension of $p$ and $p'$ an {\it adhesive extension}.

Similarly, two polymatroids $(N,h)$ and $(M,g)$ coinciding on $N\cap M$ are said to {\it adhere} 
or to have an {\it adhesive extension} if there exists a polymatroid $(N\cup M,f)$ extending 
$h,g$, i.e., $f(I)=h(I)$ for $I\subset N$, $f(J)=h(J)$ for $J\subset M$, which is also {\it 
modular} on $N$ and $M$, that is, ${f(N\cup M)=f(N)+f(M)-f(N\cap M)}$ or, equivalently, such that $N$ and $M$ are conditionally independent on the intersection $N\cap M$. As a consequence of the construction in Eq.~\eqref{eq:proext} for probabilities, restrictions of entropies have an adhesive extension, whereas general polymatroids do not (cf. Ref.~\cite{matus2007}).

This observation is at the basis of the derivation of several {\it non-Shannon} information 
inequalities, i.e., information inequalities that do not follow from 
Eqs.~\eqref{shannonineqs_basic} (cf. Ref.~\cite{matus2007}). Starting from the first non-Shannon 
inequality derived by Zhang and Yeung \cite{Zhang1998}, infinitely many inequalities have been
derived by Mat\'u\v{s} \cite{matus2007infinitely}, and several others authors investigated the problem \cite{makarychev2002new,zhang2003new,Dougherty06}.

\subsection{Marginal scenarios admitting a global extension}\label{sec:admit}

From the adhesivity property of probability distributions, one can extend probabilities defined 
on a marginal scenario to a joint probability distribution over all variables, which satisfies 
extra conditional independence  constraints that depends on the marginal scenario.

The theorem below was first  stated without proof by Vorob'ev in Ref.~\cite{vorobev1967coal}, and subsequently explicitly proven in 
Ref.~\cite{vorobev1963markov}, but also independently derived by other authors \cite{Kellerer1964a,Kellerer1964b,Malvestuto88}. The original proof, however, used a quite different terminology. It is helpful to restate it in the language of hypergraphs, and to present a sketch of it, in order to understand the role of the adhesivity property.

\begin{theorem}{\bf [Vorob'ev]}\label{th:MRFext}
A set of probabilities associated with an acyclic marginal scenario hypergraph  $\mathcal{M}$ admits a global extension to a single probability distribution. Moreover, the extension can be chosen as a MRF described by the 2-section graph $[\mathcal{M}]_2$.
\end{theorem}

{\it Sketch of the proof.--} Let $\mathcal{M}$ be the marginal scenario hypergraph, by 
definition, it is a reduced hypergraph. If $\mathcal{M}$ is acyclic, we can find an ordering 
$M_1,\ldots,M_n$ of its hyperedges respecting the running intersection property. The 
construction of a global probability distribution can then be obtained by induction on $n$, the 
number of hyperedges. For $n=1$, $P(M_1)$ is a valid probability distribution [to simplify the 
notation, we will use $P(M_i)$ as a shorthand for $P(x_{M_i})$, etc]. We then apply the 
inductive hypothesis. Let us assume that for $n-1$ $P(M_1\cup\ldots\cup M_{n-1})$ is a valid 
probability distribution extending the marginals $P(M_i)$ for $1\leq i\leq n-1$.  We want to 
extend it to $P(M_1\cup\ldots\cup M_{n-1}\cup M_n)$. By the running intersection property, 
$M_n\cap (M_1\cup\ldots\cup M_{n-1})=: S_n \subset M_j$ for $j<n$. Denoting by $P_{M_i}$ the 
marginal probability distribution on $M_i$, we define $R_n := M_n\diagdown S_n$ and
\begin{equation}
P(R_n|S_n) := \frac{P_{M_n}(M_n)}{P_{M_j}(S_n)},
\end{equation}
defining $0/0$ to be zero as in Eq.~\eqref{eq:proext}, and for the joint distribution
\begin{equation}
\begin{split}
P(M_1\cup\ldots\cup M_{n-1}\cup M_n) := P(R_n| S_n) \\P(M_1\cup\ldots\cup M_{n-1}\diagdown S_n | S_n )P(S_n).
\end{split}
\end{equation}
By the adhesivity property, this is a valid probability distribution, and its marginals 
coincide with $P(M_i)$ for $1\leq i\leq n$, so it is an extension of the marginal scenario. In 
addition, it is modular over the intersection, i.e.
\begin{equation}
(R_n \perp (M_1\cup \ldots\cup M_{n-1})\diagdown S_n | S_n)
\end{equation}
Since $\mathcal{M}$ is connected and reduced, the set of minimal separators precisely 
corresponds to the set $S_n$ above. The modularities of the constructed distribution are thus 
precisely those implied by the MRF defined by $[\mathcal{M}]_2$. $\square$

In the next section, we will see the application of this result to the general marginal scenario, i.e., not necessarily acyclic.

\subsection{Maximal set of independence conditions associated with a marginal scenario}\label{sec:max_set}

We will now see the implications of Vorob'ev's theorem on general marginal scenarios. More 
precisely, we will discuss which independence conditions, arising as MRF conditions, are 
consistent with a given marginal scenario and how to compute maximal sets of such conditions.

The main result is the following.
\begin{theorem}\label{th:extT}
Given a joint probability distribution $P$ on $n$ variables $X_1,\ldots,X_n$, and a marginal 
scenario $\mathcal{M}$, the marginals  $P_{M_i}$ for $M_i\in \mathcal{M}$ are consistent with a 
probability distribution arising from a MRF associated with the $2$-section graph 
$[\mathcal{T}]_2$,  where $\mathcal{T}$ is an acyclic hypergraph extending $\mathcal{M}$.
\end{theorem}
{\it Proof.--} An acyclic hypergraph $\mathcal{T}$ extending $\mathcal{M}$ can always be 
found. It is the clique hypergraph of a triangulation of the graph $[\mathcal{M}]_2$. The 
marginals in $\mathcal{M}$ are consistent with the marginals in $\mathcal{T}$ extracted from 
the same probability distribution $P$. Since $\mathcal{T}$ is an acyclic hypergraph, we can 
apply the construction in Th.~\ref{th:MRFext} to obtain a MRF with independence relations 
described by the 2-section graph $[\mathcal{T}]_2$. $\square$

For any given marginal scenario $\mathcal{M}$, an acyclic 
hypergraph extending it corresponds to the clique hypergraph of the triangulation of 
the 2-section $[\mathcal{M}]_2$, hence the maximum set of independence constraints will 
correspond to the triangulation with the minimum number of edges, also called the 
{\it minimum triangulation}. The problem of computing the minimum triangulation is known to be 
NP-hard \cite{Heggernes2006}. However, it is much easier to calculate  
{\it a minimal triangulation}, namely, a triangulation such that by removing any edge the 
obtained graph is no longer a chordal graph. Notice that such a minimal triangulation is not necessarily the one with the smallest number of edges among all possible triangulations. 
Several algorithms have been developed to compute a minimal triangulation, which run in $O(n+m)$ steps, $n$ being the number of nodes and $m$ the 
number of edges of a graph \cite{Heggernes2006}. 

In the following, we will adopt the above terminology also for hypergraphs, namely, we will speak about the minimum acyclic hypergraph extending $\mathcal{M}$, in the sense of the minimum triangulation, and a minimal hypergraph extending  $\mathcal{M}$, in the sense of a minimal triangulation.

\section{Optimal characterization of the marginal scenario for probabilities  and entropies}\label{sec:opt_ent}

As a consequence of the above results, the characterization of a given marginal scenario $\mathcal{M}$, in terms of inequalities for the probability vector or entropy vector, can be computed from those associated with a minimal acyclic hypergraph extending $\mathcal{M}$.
This approach offers advantages both for the probabilistic and the entropic approach. Here, we will give a brief summary of the two results, but later we mostly discuss the entropic approach.

A similar approach, albeit with a different terminology, and using a less general version of  Th.~\ref{th:extT}, has been already used in relation with Bell and noncontextuality inequalities.  For instance, the decomposition of the CHSH scenario in Fig.~\ref{fig:4a} was used in the proof of the necessity and sufficiency of Bell inequalities for the existence of a LHV model by Fine \cite{Fine1982}. Special cases of Th.~\ref{th:extT} have been discussed in Refs.~\cite{BudroniMorchio2010,budroni2012bell,Kurzy2012}, and their application to more general scenarios has been discussed for probabilities \cite{budroni2012bell2,Araujo2014} and for entropies \cite{Chaves2014}.

\subsection{Triangulation}\label{sec:tri}

The first step is to compute a minimal acyclic hypergraph $\mathcal{T}$ extending the marginal scenario $\mathcal{M}$. It can be done as follows:
\begin{itemize}
\item[a1)] Compute the 2-section graph $[\mathcal{M}]_2$.
\item[a2)] Compute its minimal triangulation.
\item[a3)] Take as $\mathcal{T}$ the corresponding clique hypergraph.
\end{itemize}

\begin{figure}[t]
    \begin{minipage}[b]{0.2\textwidth}
        \includegraphics[width=\textwidth]{sect4a_1}
         {(a) $\mathcal{M}$}
    \end{minipage}
    \hspace{0.02\textwidth}
    ~ 
    \begin{minipage}[b]{0.2\textwidth}
    \includegraphics[width=\textwidth]{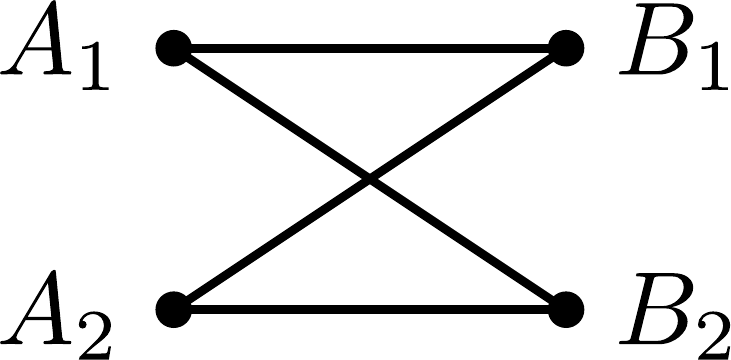}
     { (b) $[\mathcal{M}]_2$}
    \end{minipage}\\[12pt]
    \begin{minipage}[b]{0.2\textwidth}
    \includegraphics[width=\textwidth]{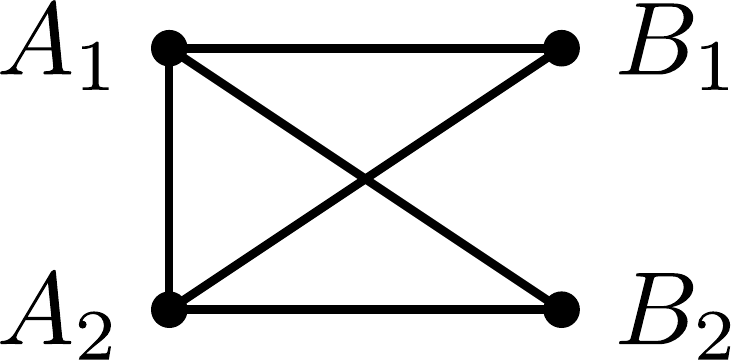}
     { (c) triangulation}
    \end{minipage}
    \hspace{0.02\textwidth}
     \begin{minipage}[b]{0.2\textwidth}
    \includegraphics[width=\textwidth]{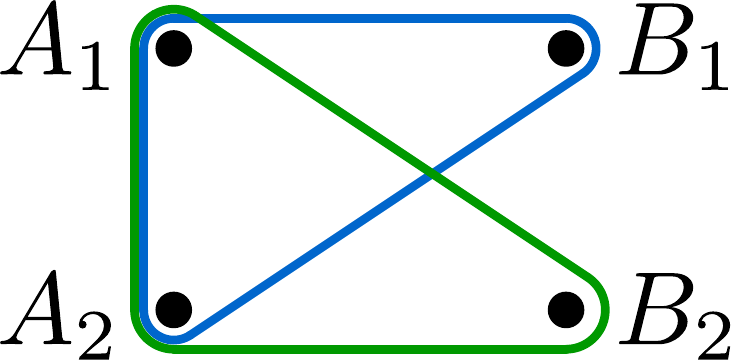}
     {(d) $\mathcal{T}$}
    \end{minipage}
    \caption{\label{fig:4a} {\bf Procedure to compute the minimal hypergraph $\mathcal{T}$ extending the marginal scenario $\mathcal{M}$ for the CHSH 
    scenario.} (a) Initial marginal scenario hypergraph corresponding to the CHSH case. (b) 2-section $[\mathcal{M}]_2$ of the original hypergraph. (c) Triangulation of the 2-section graph. (d) Clique hypergraph $\mathcal{T}$ of the triangulation. $\mathcal{T}$ extends $\mathcal{M}$. }
\end{figure}

In the following, we discuss the above procedure with a simple example. Consider the hypergraph with edges $\{ \{A_x,B_y\} \}_{x,y=1,2}$ associated with a Bell experiment and discussed in Sec.~\ref{sec:marg}. One starts with the marginal scenario hypergraph $\mathcal{M}$ of Fig.~\ref{fig:4a}~(a) and computes its 2-section  $[\mathcal{M}]_2$ [cf. Fig.~\ref{fig:4a}~(b)].  In this simple case $[\mathcal{M}]_2$ can be triangulated by adding an extra edge connecting $A_1$ and $A_2$ [cf. Fig.~\ref{fig:4a}~(c)], or equivalently, connecting $B_1$ and $B_2$. Finally, one takes as $\mathcal{T}$ the clique hypergraph of the triangulation [cf. Fig.~\ref{fig:4a}~(d)]. The corresponding MRF independence condition consistent with the marginals in $\mathcal{M}$ is $(B_1 \perp B_2 | A_1,A_2)$.

\subsection{Probabilities}
\label{sec:prob}
Once $\mathcal{T}$ has been obtained, the probabilistic inequalities describing the marginals consistent with the given scenario $\mathcal{M}$ (describing a correlation polytope, see Sec. \ref{sec:corrpoly}) can be computed as follows:
\begin{itemize}
\item[b1)] Write down the simplex inequalities associated with each maximal clique $C_i$, i.e., the inequalities corresponding to a classical probability on $|C_i|$ variables (cf. 
Ref.~\cite{Pitowsky1991} and Sec.~\ref{sec:corrpoly} for further details on the simplex polytope).
\item[b2)] Project such inequalities onto the initial marginal scenario $\mathcal{M}$ (for instance, applying the Fourier-Motzkin elimination, see Sects.~\ref{sec:e_c} and \ref{sec:corrpoly}).
\end{itemize}

\subsection{Entropies}
\label{sec:entr}

A similar approach can be applied for deriving entropic inequalities, but it only gives an outer approximation if an exact characterization of the entropy cone is not known [cf. Eq.~\eqref{eq:gammaTC} below]. 

An alternative approach can be summarized as follows:
\begin{itemize}
\item[c1)] Compute the independence constraints $\mathcal{I}(\mathcal{T})$ associated with the MRF graph $[\mathcal{T}]_2$,
\item[c2)] Consider the Shannon cone on $n$ variables $\Gamma$, with the reduced set of polymatroid axioms associated with $\mathcal{I}(\mathcal{T})$.
\item[c3)] Use the linear constraints associated with $\mathcal{I}(\mathcal{T})$, i.e., the vanishing of some conditional mutual information terms, for a partial projection of the full cone onto the marginal scenario. Then, complete the projection with the usual Fourier-Motzkin algorithm.
\end{itemize}

It is clear that the above approach can be adapted to any type of linear constraints, including those arising from some assumed causal structure, i.e., a BN or MRF, and those arising as deterministic dependence conditions, corresponding to the vanishing of some conditional entropy, discussed in Sec.~\ref{sec:c_s}.

In the next section, we will see in details what approaches are possible for characterizing entropic marginals.

\subsection{Outer approximations of the entropy cone}
\label{sec:out_app}

The adhesivity property for restrictions of an entropic polymatroid can be used to obtain outer approximations of the entropy cone as follows. 

\begin{theorem}\label{th:ad_ent_resT}
Let $\mathcal{M}$ be the marginal scenario hypergraph and $\mathcal{T}$ an acyclic hypergraph extending it. Let us denote with $\Gamma^*_\mathcal{T}$ the entropy cone intersected with the linear constraints defined by $\mathcal{I}(\mathcal{T})$. Then, we have that
\begin{equation}
\label{eq:entr_tree}
\Pi_\mathcal{M}\left(\Gamma^*\right) = \Pi_\mathcal{M}(\Gamma^*_\mathcal{T}) 
\end{equation}
where $\Pi_\mathcal{M}$ denotes the projection onto the coordinates associated with the marginal scenario $\mathcal{M}$. 
\end{theorem} 

{\it Proof.---} The result, basically, follows from the fact that the marginals in $\mathcal{M}$ are consistent with the linear constraints $\mathcal{I}(\mathcal{T})$. Given an entropic polymatroid, it is sufficient to take the associated probability distribution and apply Th.~\ref{th:extT}. The obtained distribution will be consistent with the MRF $[\mathcal{T}]_2$, hence, the associated marginal entropies will be identical to the original ones and inside $\Pi_\mathcal{M}(\Gamma^*_\mathcal{T})$ by construction.
The same construction can be combined with the limits necessary (see, e.g., Refs.~\cite{MatusLimits,Yeung2008}) to obtain the closure of entropic polymatroid $\Gamma^*$. $\square$

Inspired by the notion of adhesivity, one can also consider the outer approximation
\begin{equation}\label{eq:gammaTC}
\Gamma^*_\mathcal{T} \subset \bigcap_{k} \Gamma_{C_k}^*
\end{equation}
where $C_k$ are the hyperedges of $\mathcal{T}$ and each $\Gamma_{C_k}^*$ is the entropic cone associated with the variables in $C_k$ embedded in the space of all variables, i.e., the remaining variables are constrained. However, except for the case of the entropic polymatroid arising as a restriction of a single polymatroid and few other cases (cf. Ref.~\cite{matus2007}), it is not clear whether entropic polymatroids are adhesive; hence, the inclusion in Eq.~\eqref{eq:gammaTC} may be strict.

Now if we take an outer approximation $\Gamma$ (e.g., the Shannon cone)  of the entropic cone $\Gamma^*$, and intersect it with the linear subspaces defined by $\mathcal{I}(\mathcal{T})$, we obtain the cone $\Gamma_\mathcal{T}$ which satisfies
\begin{equation}
\label{eq:shannon_tree}
\Pi_\mathcal{M}\left(\Gamma^*\right) =  \Pi_\mathcal{M}\left(\Gamma^*_\mathcal{T}\right) \subset \Pi_\mathcal{M}\left(\Gamma_\mathcal{T}\right) \subset \Pi_\mathcal{M}\left(\bigcap_{k} \Gamma_{C_k}\right).
\end{equation}

In general, given a marginal scenario $\mathcal{M}$ there exist several minimal acyclic hypergraphs $\{ \mathcal{T}_i \}$ extending it. The inclusion in Eq.~\eqref{eq:shannon_tree} is valid for each $\Pi_\mathcal{M}\left(\Gamma_{\mathcal{T}_i}\right)$, consequently, also for the intersection
\begin{equation}
\label{eq:shannon_tree_int}
\Pi_\mathcal{M}\left(\Gamma^*\right) \subset \bigcap_i \Pi_\mathcal{M}\left(\Gamma_{\mathcal{T}_i}\right).
\end{equation}

As a conclusion, for each marginal scenario $\mathcal{M}$, we have three different outer approximations for $\Pi_\mathcal{M}\left(\Gamma^*\right)$, namely,
\begin{itemize}
\item[(i)] the intersection of the projections of the full Shannon cones $\Gamma_{[n]}$ with modularity conditions $\{\mathcal{I}(\mathcal{T}_i)\}$, namely, $\bigcap_{i} \Pi_\mathcal{M}\left(\Gamma_{\mathcal{T}_i}\right)$;  $\{\mathcal{T}_i\}$ is the set of minimal acyclic hypergraphs extending $\mathcal{M}$;
\item[(ii)] the projection of the full Shannon cone, namely,  $\Pi_\mathcal{M}\left(\Gamma\right)$;
\item[(iii)] the projection of the intersection of Shannon cones associated with $\{C_k^{(i)}\}_k$ is the set of hyperedges of $\mathcal{T}_i$, i.e., $\bigcap_{i} \Pi_\mathcal{M}\left(\bigcap_{k} \Gamma_{C_k^{(i)}}\right)$, where each $\Gamma_{C_k^{(i)}}$ is seen as a cone in the space of all variables, and the variables not appearing in $C_k^{(i)}$ are unconstrained [cf. Eq.~\ref{eq:gammaTC}].
\end{itemize}

We can then summarize the relations among the above cones as follows
\begin{obs}
\label{obs:cones_inclusion}
The above approximations satisfy the inclusion relations
\begin{equation}
\Pi_\mathcal{M}\left(\Gamma^*\right)\subset \bigcap_{i} \Pi_\mathcal{M}\left(\Gamma_{\mathcal{T}_i}\right) \subset \Pi_\mathcal{M}\left(\Gamma\right) \subset \bigcap_{i} \Pi_\mathcal{M}\left(\bigcap_{k} \Gamma_{C_k^{(i)}}\right).
\end{equation}
\end{obs}

In general, the inclusion relations from Observation \ref{obs:cones_inclusion} are proper.  In particular, it means that the outer approximation $\bigcap_{i\in\{\mathcal{T}_i\} }\Pi_\mathcal{M}\left(\Gamma_{\mathcal{T}_i}\right)$ is tighter than the projection of the Shannon cone ---the most widely used method in the literature; see for instance \cite{Zhang1998,matus2007,matus2007infinitely,Yeung2008}--- and thus may contain nonconstrained non-Shannon-type inequalities. We will provide some examples of this in Sec.~\ref{sec:ex}.

\section{Indistinguishability of causal structures}
\label{sec:ind}

In this section, we will investigate the role of the above results for the case of probability distributions and entropies, where some underlying causal structure is assumed, i.e.,  some additional conditional independence constraints are present. 

The general goal is to characterize the region of probability distributions or entropies compatible with a given causal structure. Via such a characterization, for instance via Bell inequalities, one can check whether some observed data are consistent or inconsistent with the assumed causal structure, thus being of fundamental importance in both quantum information and any other field where causal discovery may play a relevant role. Furthermore, notice that unless one is able to intervene in the physical system under investigation \cite{Pearlbook}, one can never unambiguously prove what is the underlying causal structure. Rather, based on observations alone, one can only prove the compatibility or incompatibility of a given presumed set of causal relations. As expected, the less constraints a given causal structure implies on the distributions compatible with it, the more correlations such models can explain and the smaller is the possibility of falsifying it.

The ideas to be discussed next apply not only to the case of classical causal structures, but also to the quantum \cite{Fritz2012,Leifer2013,chaves2014information,Pienaar2014,fritz2016beyond,Costa2016} and even post-quantum \cite{Henson2014,Chaves2016entropic,pienaar2016causal} generalizations as well. In the following, we will focus on the classical case, that is, all nodes in the associated Bayesian networks or MRFs represent random variables for which a global joint probability distribution can always be assumed to exist. The case of quantum and post-quantum theories will be briefly discussed at the end of this section and presented in full details elsewhere. 

Let us consider the causal structure defined by a graph $\mathcal{G}$, which may be either a DAG corresponding to a Bayesian network, or a graph corresponding to a MRF. We will denote the set of independence conditions associated with $\mathcal{G}$ as $\mathcal{I}(\mathcal{G})$, as in Eqs.\eqref{eq:bayes_ind},\eqref{eq:mrf_ind}. Let us now assume to have a fixed marginal scenario, with $\mathcal{M}$ the associated hypergraph. Let $\{\mathcal{T}_i\}_i$ be the set of acyclic hypergraph extending $\mathcal{M}$ as in Th.~\ref{th:MRFext}. We will denote by $\mathcal{I}(\mathcal{T}_i)$, the set of independence conditions of the corresponding MRF defined by its 2-section $[\mathcal{T}_i]_2$. We have the following result

\begin{theorem}
\label{th:caus_and_adh}
Given $\mathcal{M},\mathcal{G}$ and $\{\mathcal{T}_i\}_i$, we have three possible cases:
\begin{itemize}
\item[$(i)$]  $\exists i$ such that $\mathcal{I}(\mathcal{G})\subset\mathcal{I}(\mathcal{T}_i).$
\item[$(ii)$] $\forall i$ $\mathcal{I}(\mathcal{T}_i)\subset \mathcal{I}(\mathcal{G}).$
\item[$(iii)$] $\forall i$  $\mathcal{I}(\mathcal{G})\not \subset \mathcal{I}(\mathcal{T}_i)$, and $\exists j$ such that $\mathcal{I}(\mathcal{T}_j)\not \subset\mathcal{I}(\mathcal{G})$.
\end{itemize}
Then:

In case $(i)$, it is impossible to falsify the causal structure described by $\mathcal{G}$. This follows since for any probability distribution $P$, its marginals in $\mathcal{M}$ are always consistent with the causal structure described by $\mathcal{G}$. Approach (c1)--(c3) of Sec.~\ref{sec:opt_ent} can be used to characterize the marginals associated with $\mathcal{M}$ and $\mathcal{G}$. 

In case $(ii)$, marginals associated with $\mathcal{M}$ can still generate correlations that are incompatible with the causal structure associated with $\mathcal{I}(\mathcal{G})$. Approach (c1)--(c3) can be used, but the obtained constraints are redundant with respect to $\mathcal{I}(\mathcal{G})$. It is, therefore, more convenient to apply approach (c1)--(c3) directly with the conditional independence relations  $\mathcal{I}(\mathcal{G})$. 

In case $(iii)$, the marginal correlations associated with $\mathcal{M}$ can again be incompatible with the causal structure associated with $\mathcal{I}(\mathcal{G})$. However, the marginal scenario implies constraints that cannot be combined with those of the causal structure $\mathcal{G}$. Hence, the approach (c1)-(c3) cannot be used with the constraints $\mathcal{I}(\mathcal{T})$.
\end{theorem}

{\it Proof.---} In case $(i)$, the independence constraints of the causal structure are just a subset of the independence constraints consistent with the marginal scenario. As a consequence, given the marginal probabilities $\{P_M\}_{M\in\mathcal{M}}$, one can repeat the construction of Th.~\ref{th:MRFext} and obtain a valid joint probability distribution $P$ that is consistent with the causal structure defined by $\mathcal{G}$. 

The above result implies that the approach of Sec.~\ref{sec:opt_ent}, both the constructions (b1)--(b3) for probabilities and (c1)--(c3) for entropies, applies also to the case of a causal structure with $\mathcal{I}(\mathcal{G})\subset\mathcal{I}(\mathcal{T}_i)$. 

In case $(ii)$, the marginal scenario implies fewer constraints than the causal structure $\mathcal{G}$, hence, it is clear that the marginals associated with $\mathcal{M}$ are still able to detect inconsistencies with the causal structure associated with $\mathcal{I}(\mathcal{G})$. 

From the point of view of the characterization, one can still use the approach (c1)--(c3) of Sec.~\ref{sec:opt_ent}. However, it is more convenient to use in (c3) the linear constraints implied by $\mathcal{I}(\mathcal{G})$, since they also include those associated with $\mathcal{I}(\mathcal{T}_i)$ for all $i$.  

In case $(iii)$, it is again clear that the marginals associated with $\mathcal{M}$ are still 
able to detect inconsistencies with the causal structure associated with 
$\mathcal{I}(\mathcal{G})$. However, the approach (c1)--(c3) of Sec.~\ref{sec:opt_ent} cannot 
be used as it generates constraints inconsistent with the causal structure. The situation is 
clarified by the following inclusion relations among entropy cones in 
Eqs.~\eqref{eq:caus_cons1}--\eqref{eq:caus_cons3}.
Let us denote by $L_{\mathcal{G}}$ the subspace of entropy vectors in $\mathbb{R}^{2^n}$ where
the linear constraints imposed by $\mathcal{I}(\mathcal{G})$ are satisfied, and similarly for
$L_{\mathcal{T}_i}$. The entropy cone associated with a given causal structure, either
$\mathcal{G}$ or $\mathcal{T}_i$, will be $\Gamma^*\cap L_{\mathcal{G},\mathcal{T}_i}$. We can 
write down the relation between the associated entropy cones as
\begin{eqnarray}\label{eq:caus_cons1}
\Pi_\mathcal{M}\left(\Gamma^*\right) &=& \Pi_\mathcal{M}(\Gamma^*\cap L_{\mathcal{T}_i}), \\
\Pi_\mathcal{M}\left(\Gamma^*\cap L_{\mathcal{G}}\cap L_{\mathcal{T}_i}\right) &\subset& \Pi_\mathcal{M}(\Gamma^*\cap L_{\mathcal{G}}),\label{eq:caus_cons2}\\
\label{eq:caus_cons3} \Pi_\mathcal{M}\left(\Gamma^*\cap L_{\mathcal{G}}\right) &\subset& \Pi_\mathcal{M}(\Gamma^*) \cap \Pi_\mathcal{M}( L_{\mathcal{G}})\\
\nonumber &=& \Pi_\mathcal{M}(\Gamma^*\cap L_{\mathcal{T}_i})\cap \Pi_\mathcal{M}( L_{\mathcal{G}})
\end{eqnarray}
It is then clear that by imposing both the $\mathcal{I}(\mathcal{G})$ and $\mathcal{I}(\mathcal{T}_i)$ conditions one obtains an inner approximation of the entropy cone associated with the causal structure. Vice versa, imposing the causal structure conditions after the projection gives an outer approximation.$\square$

To clarify this last part, in particular the impossibility of combining the independence relations $\mathcal{I}(\mathcal{G})$ arising from the causal structure, with the $\{ \mathcal{I}(\mathcal{T}_i)\}_i$ arising from the marginal scenario, it is helpful to look at some specific examples. We will discuss them in details in Sec.~\ref{sec:3cas}. 

A natural question is, then, how to extend the above results to the case of quantum and post-quantum causal structures. For instance, in the postquantum case if $\mathcal{M}$ is an acyclic hypergraph, then by Th. \ref{th:MRFext} the observed marginals are always consistent with a classical probability distribution. The same holds, in particular, in the quantum case. However, the fact that different rules for causal inference arise in the quantum and postquantum cases (cf. \cite{chaves2014information,Henson2014,Pienaar2014,Costa2016,Chaves2016entropic,pienaar2016causal}) together with the different characterization of the associated entropy regions (cf. \cite{chaves2014information, Chaves2016entropic}) makes the above investigation more complex and worth a separate discussion elsewhere \cite{inprep}.

\section{Examples and computational results}
\label{sec:ex}
In order to clarify the results and methods presented in the previous sections, we discuss examples of marginal scenarios and causal structures, together with some computational results.  
\subsection{Inclusions in Obs.~\ref{obs:cones_inclusion}}
In the following, we will discuss the possible cases presented in Obs.~\ref{obs:cones_inclusion}. In particular, we will see examples of strict and non-strict inclusion for the outer approximations of the entropy cone.
\subsubsection{Case: $\bigcap_{i}\Pi_\mathcal{M}\left(\Gamma_{\mathcal{T}_i}\right) \subsetneq \Pi_\mathcal{M}\left(\Gamma\right) \subsetneq \bigcap_{i} \Pi_\mathcal{M}\left(\bigcap_{k} \Gamma_{C_k^{(i)}}\right)$}

\begin{figure}[t]
\vspace{12pt}
    \begin{minipage}[b]{0.2\textwidth}
        \includegraphics[width=\textwidth]{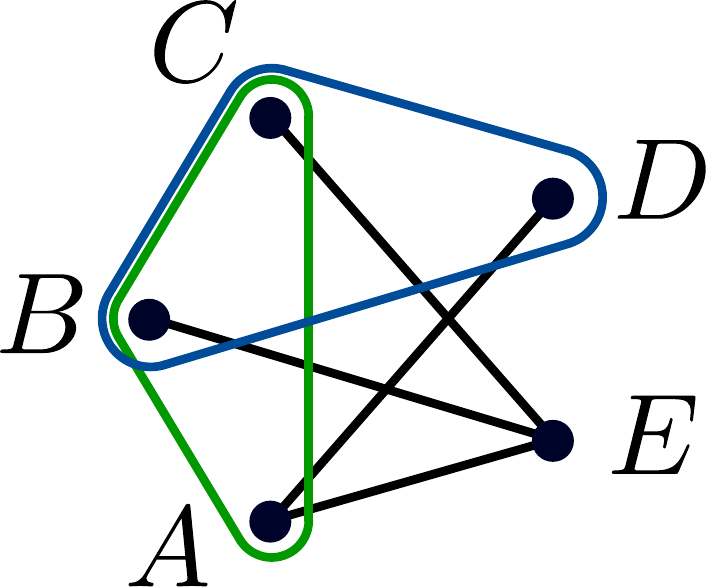}
         { (a) }
    \end{minipage}
    \hspace{0.02\textwidth}
    ~ 
    \begin{minipage}[b]{0.2\textwidth}
    \includegraphics[width=\textwidth]{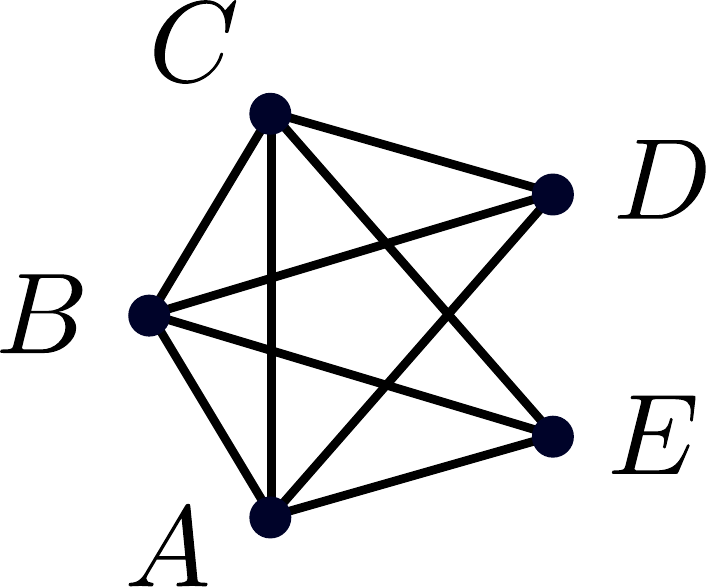}
     { (b) }
    \end{minipage}
    \caption{\label{fig:6a} Marginal scenario hypergraph (a) and (b) its 2-section}
\end{figure}

As already noted by Mat\'{u}\v{s} \cite{matus2007}, the proper inclusion $\bigcap_{i}\Pi_\mathcal{M}\left(\Gamma_{\mathcal{T}_i}\right) \subsetneq \Pi_\mathcal{M}\left(\Gamma\right)$ means existence of non-Shannon-type inequalities in  $\Pi_\mathcal{M}\left(\Gamma_{\mathcal{T}_i}\right)$. On the other hand, the strict inclusion $\Pi_\mathcal{M}\left(\Gamma\right) \subsetneq \bigcap_{i} \Pi_\mathcal{M}\left(\bigcap_{k} \Gamma_{C_k^{(i)}}\right)$ is related to the non-adhesivity of general polymatroids.

To construct an example, is necessary to take at least four variables ~\cite{matus2007}, and our example will consist of five variables. Let us consider the following marginal scenario $\mathcal{M} = \{ABC, BCD, AE, BE, CE, AD\}$, shown on the Fig.~\ref{fig:6a}~(a). One can easily see that the 2-section of the hypergraph $\mathcal{M}$ is a triangulated graph shown in  Fig.~\ref{fig:6a}~(b) and thus there is only one corresponding clique hypergraph $\mathcal{T} = \{ABCD, ABCE\}$. The independence constraint $I(D:E|ABC) = 0$ arising from the adhesivity property, gives rise, after projection on the set of entropies given by $\left\{ B,C,D,AD,AE,BD,BE,CD,CE,ABC,BCD \right\}$, to the following 3 non-redundant non-Shannon-type inequalities
\begin{eqnarray}
& & H(E|C) + H(C|D) + H(E|B) + H(B,D) + H(A,D) \nonumber \\
&-& s_1H(B,C,D) - s_2H(A,E) + s_3H(A,B,C) \geq 0,
\label{eq:nonshineq}
\end{eqnarray}
where the coefficient triplet $(s_1,s_2,s_3) \in \{(1,2,1), (2,1,1), (2,2,2)\}$.

Another interesting aspect of this example is the reduction in the computational time required to compute the projection on a usual desktop computer. More precisely, adding the linear constraint $I(D:E|ABC) = 0$ reduced the time of our computation for the projection from approximately $320$ to only $27$ seconds.

\subsubsection{Case: $\bigcap_{i}\Pi_\mathcal{M}\left(\Gamma_{\mathcal{T}_i}\right) = \Pi_\mathcal{M}\left(\Gamma\right) = \bigcap_{i} \Pi_\mathcal{M}\left(\bigcap_{k} \Gamma_{C_k^{(i)}}\right)$}

\begin{figure}[t]
\vspace{12pt}
    \begin{minipage}[b]{0.2\textwidth}
        \includegraphics[width=\textwidth]{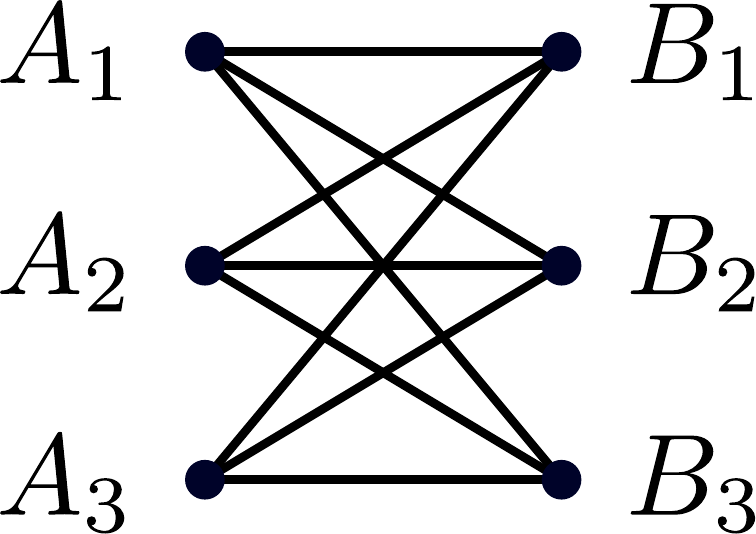}
         { (a) }
    \end{minipage}
    \hspace{0.02\textwidth}
    \begin{minipage}[b]{0.2\textwidth}
    \includegraphics[width=\textwidth]{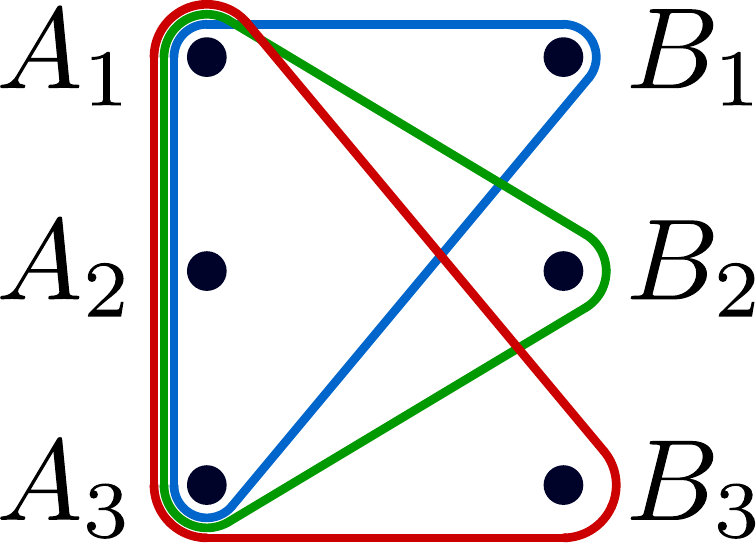}
     { (b) }
    \end{minipage}
    \caption{\label{fig:6b} Marginal scenario (a) and a clique hypergraph (b)}
\end{figure}

Consider the marginal scenario $\mathcal{M} = \{A_iB_j\}, \forall i,j \in \{1,2,3\}$, shown in Fig.~\ref{fig:6b}~(a), corresponding to a bipartite Bell scenario with three measurement settings per party. The clique hypergraph of one of the triangulations of $\mathcal{M}$ is shown on Fig.~\ref{fig:6b}~(b).

If we consider an intersection of cones $\Gamma_{C_k^{(1)}}$ for cliques $C^{(1)}_k = \{A_1,A_2,A_3,B_k \}$, $k=1,2,3$, which are the edges of the hypergraph shown on Fig.~\ref{fig:6b}~(b), we find that its projection on the marginal scenario $\mathcal{M}$ differs from the projection of the full cone such that 108 out of 217 rays of the projection $\Pi_\mathcal{M}\left(\bigcap_{k} \Gamma_{C_k^{(1)}}\right)$ are outside of $\Pi_\mathcal{M}\left(\Gamma\right)$. This can be checked via linear programming (cf. Ref.~\cite{Chaves2016entropic} Sec.~II of the Supplemental Material) by simply checking whether such rays are compatible with the basic Shannon inequalities characterizing $\Gamma$.

However, if we now consider cliques of the second possible triangulation of $\mathcal{M}$, which are $C^{(2)}_k = \{A_k,B_1,B_2,B_3 \}$, $k=1,2,3$ and compute the intersection $\bigcap_{i=1,2} \Pi_\mathcal{M}\left(\bigcap_{k} \Gamma_{C_k^{(i)}}\right)$ we find, again via linear programming, that all its extremal rays are inside, not only $\Pi_\mathcal{M}\left(\Gamma\right)$, but also $\Pi_\mathcal{M}\left(\Gamma_{\mathcal{T}_i}\right)$, for all $i$. Hence we have that all the outer approximations coincide. 

The reasons why such an equivalence is interesting is that the calculation of $\bigcap_{i=1,2} \Pi_\mathcal{M}\left(\bigcap_{k} \Gamma_{C_k^{(i)}}\right)$ with a standard Fourier-Motzkin algorithm on a standard desktop takes few minutes, however, a direct computation of $\Pi_\mathcal{M}\left(\Gamma\right)$ or $\Pi_\mathcal{M}\left(\Gamma_{\mathcal{T}_i}\right)$ seems to be out of computational reach (at least on a usual desktop computer). 

\subsection{Three cases in Theorem \ref{th:caus_and_adh}}\label{sec:3cas}
 
In this subsection, we will discuss in detail and provide examples for the different cases presented in Theorem \ref{th:caus_and_adh}.

\begin{figure}[t]
    \begin{minipage}[b]{0.2\textwidth}
        \includegraphics[width=\textwidth]{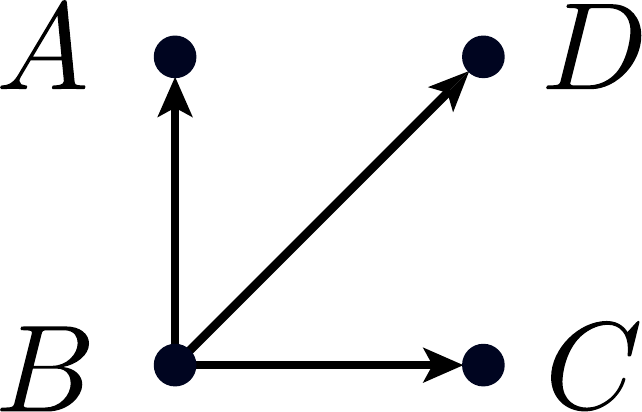}
         { (a) $\mathcal{G}_1$}
    \end{minipage}
    \hspace{0.02\textwidth}
    \begin{minipage}[b]{0.2\textwidth}
    \includegraphics[width=\textwidth]{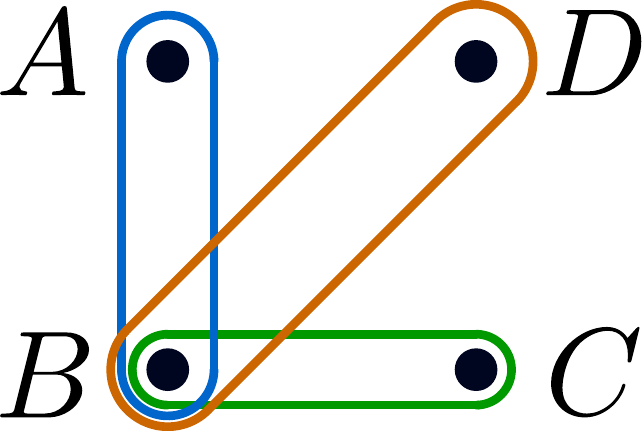}
     { (b) $\mathcal{M}_1$}
    \end{minipage}\\[12pt]
    \begin{minipage}[b]{0.2\textwidth}
    \includegraphics[width=\textwidth]{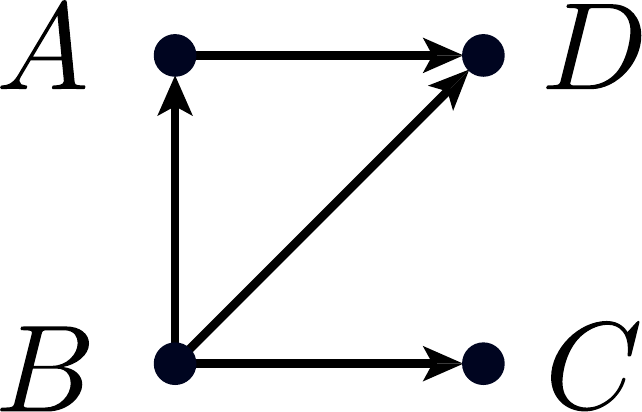}
     {(c) $\mathcal{G}_2$}
    \end{minipage}
    \hspace{0.02\textwidth}
     \begin{minipage}[b]{0.2\textwidth}
    \includegraphics[width=\textwidth]{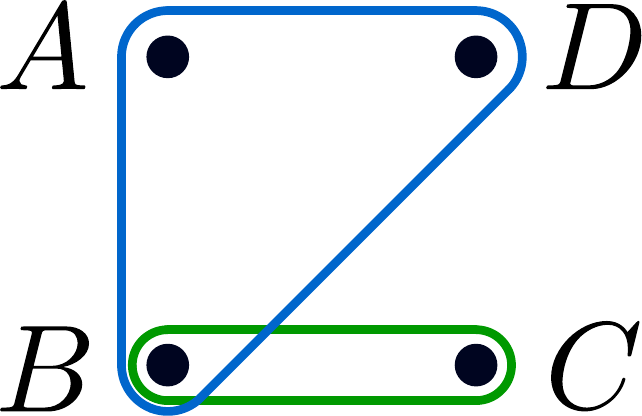}
     {(d) $\mathcal{M}_2$}
    \end{minipage}
    \caption{\label{fig:6c} Example of causal structures (a),(c), and marginal scenarios (b),(d).}
\end{figure}

Let us consider four random variables $A,B,C,D$. 

\subsubsection{Case (i): $\exists i$ such that $\mathcal{I}(\mathcal{G})\subset\mathcal{I}(\mathcal{T}_i)$}

Let us consider the marginal scenario given by $\mathcal{M}_1 = \{AB,BD,BC\}$ and shown in Fig.~\ref{fig:6c}~(b). One can easily see that the clique hypergraph $\mathcal{T}$ of the corresponding triangulation of the 2-section graph is unique and coincides with $\mathcal{M}_1$. The independence constraints implied by $\mathcal{T}$ are given by
\begin{eqnarray}
(A\perp C \,|\, B) \nonumber \\
(C\perp D \,|\, B) \label{eq:th3m1}\\
(D\perp A \,|\, B) \nonumber
\end{eqnarray}   

Consider now the two causal structures $\mathcal{G}_1$ and $\mathcal{G}_2$ shown in Fig.~\ref{fig:6c}~(a),(c). The corresponding independence constrains are 
$\{(A\perp C \,|\, B\,); (C\perp D \,|\, B\,); (D\perp A \,|\, B\,)\}$ for $\mathcal{G}_1$ and $\{(A\perp C \,|\, B\,); (C\perp D \,|\, B\,)\}$ for $\mathcal{G}_2$. In both cases $\mathcal{I}(\mathcal{G}_{1,2})\subset\mathcal{I}(\mathcal{T})$ which means that marginal scenario $\mathcal{M}_1$ from Fig.~\ref{fig:6c}~(b) is insufficient to distinguish causal structures $\mathcal{G}_1$ and $\mathcal{G}_2$. In turn, a marginal scenario, which would be enough to distinguish between these two causal structures is given by $\mathcal{M}_2 = \{ABD,BC\}$ and shown in Fig.~\ref{fig:6c}~(d). 

\subsubsection{Case (ii): $\forall i$ $\mathcal{I}(\mathcal{T}_i)\subset \mathcal{I}(\mathcal{G})$}

Consider again the causal graph $\mathcal{G}_1$. As we already noted the independence constraints associated with this graph are 
\begin{eqnarray}
(A\perp C \,|\, B) \nonumber \\
(C\perp D \,|\, B) \label{eq:th3m2}\\
(D\perp A \,|\, B) \nonumber
\end{eqnarray} 
If we are now interested in the marginal scenario $\mathcal{M}_2 = \{ABD,BC\}$, then one can see that in that case there is again only one possible triangulation of the 2-section graph of $\mathcal{M}_2$ and consequently only one corresponding clique hypergraph. The set of independence constraints, consistent with $\mathcal{M}_2$ is $\{(A\perp C \,|\, B\,); (C\perp D \,|\, B\,)\}$, which is a subset of constraints from Eq.~\eqref{eq:th3m2}. In other words, constraints coming from marginal scenario $\mathcal{M}_2$ are redundant to those coming from the causal structure.

\subsubsection{Case (iii): $\forall i$  $\mathcal{I}(\mathcal{G})\not \subset \mathcal{I}(\mathcal{T}_i)$, and $\exists j$ such that $\mathcal{I}(\mathcal{T}_j)\not \subset\mathcal{I}(\mathcal{G})$}

The third one is, arguably, the most interesting case: it shows that the independence constraints arising from the marginal scenario may be ``inconsistent'' with those associated with the causal structure. An example where this problem arises is the classical case of the information causality scenario \cite{Pawlowski2009}: Alice receives two independent inputs $X_0,X_1$, she creates a message $M$ depending on those inputs that is sent to Bob who provide guesses $Y_0,Y_1$, respectively of $X_0,X_1$, on the basis of the message $M$.

\begin{figure}[t]
\vspace{12pt}
    \begin{minipage}[b]{0.2\textwidth}
        \includegraphics[width=\textwidth]{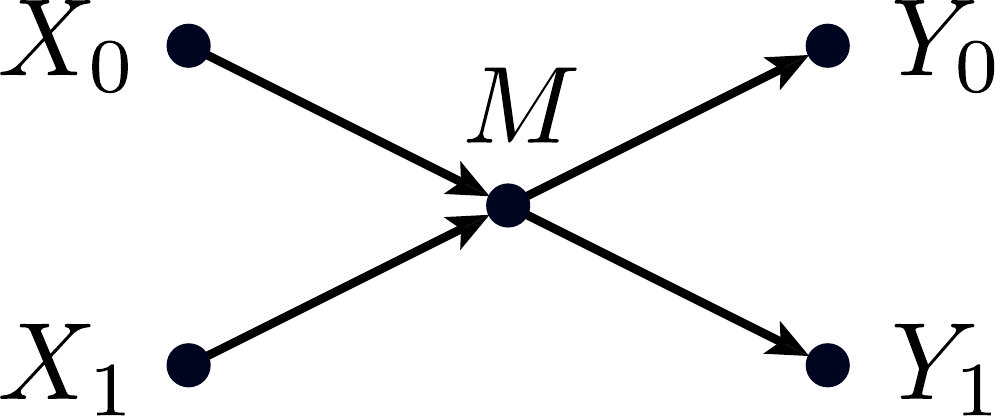}
         { (a) }
    \end{minipage}
    \hspace{0.02\textwidth}
    \begin{minipage}[b]{0.2\textwidth}
    \includegraphics[width=\textwidth]{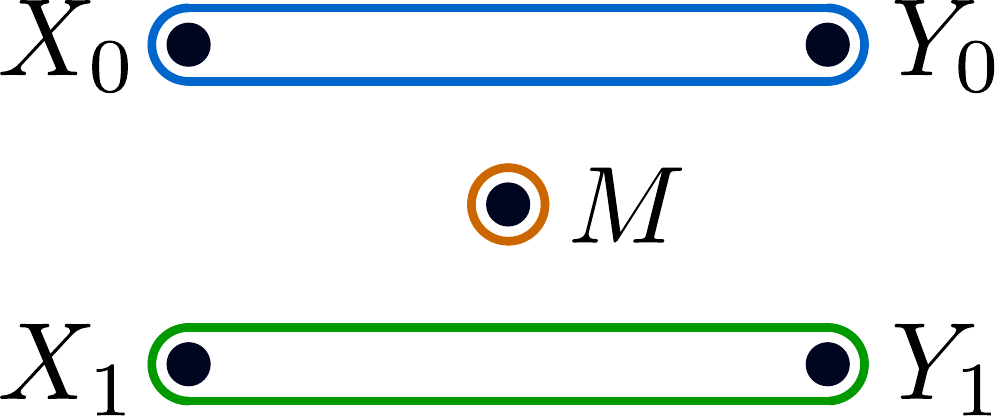}
     { (b) }
    \end{minipage}
    \caption{\label{fig:6d} Causal structure (a) and marginal scenario (b) of classical case of information causality. $X_0,X_1$ are random inputs for Alice, $Y_0,Y_1$ are guesses for Bob, and $M$ is a message which Alice sends to Bob.}
\end{figure}

The corresponding causal structure is  shown in Fig.~\ref{fig:6d}~(a) and the marginal scenario $\mathcal{M}$ in Fig.~\ref{fig:6d}~(b). Once again the clique hypergraph $\mathcal{T}$ coincides with $\mathcal{M}$, hence, it is unique. 

We need to show that
\begin{equation}
\mathcal{I}(\mathcal{G})\not \subset \mathcal{I}(\mathcal{T}),\; \text{and}\; \mathcal{I}(\mathcal{T})\not \subset\mathcal{I}(\mathcal{G}).
\label{eq:case3th3}
\end{equation}
For showing $\mathcal{I}(\mathcal{G}) \not \subset \mathcal{I}(\mathcal{T})$, we consider the conditional independence between inputs $X_0,X_1$ and guesses $Y_0,Y_1$. That is,
\begin{equation}
\{X_i\perp Y_j\,|\, M\}_{i,j=0,1} \not \subset \mathcal{I}(\mathcal{T}).
\end{equation}
An example in the other direction is an independence of message $M$ from the rest of the variables, which is implied by $\mathcal{T}$, and is not consistent with conditional independences $\mathcal{I}(\mathcal{G})$.

The projection $\Pi_\mathcal{M}(\Gamma\cap L_{\mathcal{G}})$ gives rise to the following inequalities
\begin{subequations}
\begin{align}
I(X_0:Y_0) \geq 0, \quad I(X_1:Y_1) \geq 0, \label{eq:infcaus1} \\
H(Y_0|X_0) \geq 0, \quad H(X_0|Y_0) \geq 0, \label{eq:infcaus2}\\
H(Y_1|X_1) \geq 0, \quad H(X_1|Y_1) \geq 0, \label{eq:infcaus3}\\
I(X_0:Y_0) + I(X_1:Y_1) \leq H(M) \label{eq:infcaus4},
\end{align}\label{eq:infcaus}
\end{subequations}

\noindent{}where inequalities \eqref{eq:infcaus1}, 
\eqref{eq:infcaus2}, and \eqref{eq:infcaus3} are simply polymatroid axioms for the marginals 
$\{X_0Y_0,X_1Y_1\}$ and one obtains these 6 inequalities, if one computes 
$\Pi_\mathcal{M}\left(\Gamma\cap L_{\mathcal{T}}\right)$. The last inequality 
Eq.~(\ref{eq:infcaus4}) is the information causality inequality and is not implied by 
$\mathcal{I}(\mathcal{T})$.   

As a result of the relation from Eq.~(\ref{eq:case3th3}), one cannot combine $\mathcal{I}
(\mathcal{T})$ and $\mathcal{I}(\mathcal{G})$. Due to the relation from 
Eq.~(\ref{eq:caus_cons2}) and the fact that the Shannon cone is an outer approximation for the 
case of more than 3 variables, the projection of the Shannon cone with combined constraints 
$\mathcal{I}(\mathcal{T})$ and $\mathcal{I}(\mathcal{G})$ in this case provides neither an 
outer nor an inner approximation of $\Pi_\mathcal{M}\left(\Gamma^*\cap L_{\mathcal{G}}\right)$.

\section{Conclusion}
\label{sec:concl}

Deciding global features of a system of interest with limited information, the so-called marginal problem, is a task often encountered in many fundamental and practical problems. In turn, causal discovery, the inference of causal relations underlying the correlations between observed variables, is yet another basic goal in the most diverse fields. In this paper, we use the notion of adhesivity to investigate marginal problems within causal inference. In particular, we show which causal relations are always compatible with some given marginal information. As a consequence, we are able to identify which causal structures, describing either a Bayesian network or a Markov random field, can be distinguished when only limited marginals are available. In addition, our results provide a method for a faster characterization (in terms of Bell inequalities) of the marginal scenarios associated with a given causal model. This holds true for the both the probabilistic and entropic approaches for Bell inequalities. In particular, in the entropic case our construction allows for a more accurate characterization of allowed regions for entropic marginals, as shown with explicit computational results. 

An immediate and interesting open question is the possible generalization of these results to the case where the causal relations between the variables are mediated via quantum or postquantum (non-signalling) resources. Quantum generalizations of the notion of a causal structure have attracted growing attention \cite{Leifer2013,fritz2016beyond,Henson2014,chaves2014information,Pienaar2014,Costa2016} and we believe that our results could constitute a viable option for the characterization of such quantum structures. Partial results, such as the fact that classical and postquantum correlations coincide for the case of acyclic marginal scenario hypergraphs (cf. Sec.~\ref{sec:ind}), show that a similar approach can be extended also to the quantum and postquantum cases. In particular, this investigation could lead to new insights on which causal structures can demonstrate some sort of non-locality \cite{Henson2014,pienaar2016causal}. Finally, another possibility is to try to combine the notion of adhesivity and the algebraic geometry tools \cite{Geiger1999} required to characterize the set of compatible probabilities associated with complex causal structures \cite{lee2015causal,Chaves2016poly,Rosset2016}.

\begin{acknowledgments}
CB and NM acknowledge financial support from the DFG, the ERC (Consolidator Grant 683107/TempoQ), the FQXi Fund (Silicon Valley Community Foundation) and the DAAD.
RC acknowledges financial support from the Brazilian ministries MEC and MCTIC, the FQXi Fund, the Excellence Initiative of the German Federal and State Governments (Grants ZUK 43 and 81), the US Army Research Office under Contracts No. W911NF-14-1-0098 and No. W911NF-14-1-0133 (Quantum Characterization, Verification, and Validation), and the DFG (GRO 4334 and SPP 1798). 
\end{acknowledgments}



\bibliography{entropic_ineq}

\end{document}